\documentclass[onecolumn,authoryear]{els-mrw} 

\usepackage{amsmath,amssymb,amsfonts,amsthm,makeidx,graphicx}
\usepackage{txfonts}
\usepackage{helvet}
\usepackage{rotating}

\def\DeclareAbbreviation#1#2{%
   \DeclareRobustCommand*#1{\@journalname{#2}}}
\def\@journalname#1{{\normalfont#1}}
\DeclareAbbreviation\aj{AJ}
\DeclareAbbreviation\araa{ARA\&A}
\DeclareAbbreviation\apj{ApJ}
\DeclareAbbreviation\apjl{ApJL}
\DeclareAbbreviation\apjs{ApJS}
\DeclareAbbreviation\ao{Appl.\ Opt.}
\DeclareAbbreviation\apss{Ap\&SS}
\DeclareAbbreviation\aap{A\&A}
\DeclareAbbreviation\aapr{A\&AR}
\DeclareAbbreviation\aaps{A\&AS}
\DeclareAbbreviation\azh{AZh}
\DeclareAbbreviation\baas{BAAS}
\DeclareAbbreviation\jrasc{JRASC}
\DeclareAbbreviation\memras{MmRAS}
\DeclareAbbreviation\mnras{MNRAS}
\DeclareAbbreviation\pra{Phys.\ Rev.\ A}
\DeclareAbbreviation\prb{Phys.\ Rev.\ B}
\DeclareAbbreviation\prc{Phys.\ Rev.\ C}
\DeclareAbbreviation\prd{Phys.\ Rev.\ D}
\DeclareAbbreviation\pre{Phys.\ Rev.\ E}
\DeclareAbbreviation\prl{Phys.\ Rev.\ Lett.}
\DeclareAbbreviation\pasp{PASP}
\DeclareAbbreviation\pasj{PASJ}
\DeclareAbbreviation\pasa{PASA}
\DeclareAbbreviation\qjras{QJRAS}
\DeclareAbbreviation\raa{Res.\ Astron.\ Astronphys.}
\DeclareAbbreviation\skytel{S\&T}
\DeclareAbbreviation\solphys{Sol.\ Phys.}
\DeclareAbbreviation\sovast{Soviet\ Ast.}
\DeclareAbbreviation\ssr{Space\ Sci.\ Rev.}
\DeclareAbbreviation\zap{ZAp}
\DeclareAbbreviation\nat{Nature}
\DeclareAbbreviation\na{New Astron.}
\DeclareAbbreviation\nar{New Astron. Rev.}
\DeclareAbbreviation\iaucirc{IAU\ Circ.}
\DeclareAbbreviation\aplett{Astrophys.\ Lett.}
\DeclareAbbreviation\apspr{Astrophys.\ Space\ Phys.\ Res.}
\DeclareAbbreviation\bain{Bull.\ Astron.\ Inst.\ Netherlands}
\DeclareAbbreviation\fcp{Fund.\ Cosmic\ Phys.}
\DeclareAbbreviation\gca{Geochim.\ Cosmochim.\ Acta}
\DeclareAbbreviation\grl{Geophys.\ Res.\ Lett.}
\DeclareAbbreviation\jcp{J.\ Chem.\ Phys.}
\DeclareAbbreviation\jgr{J.\ Geophys.\ Res.}
\DeclareAbbreviation\jqsrt{J.\ Quant.\ Spectrosc.\ Radiat.\ Transfer}
\DeclareAbbreviation\memsai{Mem.\ Soc.\ Astron.\ Italiana}
\DeclareAbbreviation\nphysa{Nucl.\ Phys.\ A}
\DeclareAbbreviation\physrep{Phys.\ Rep.}
\DeclareAbbreviation\physscr{Phys.\ Scr.}
\DeclareAbbreviation\planss{Planet.\ Space\ Sci.}
\DeclareAbbreviation\procspie{Proc.\ SPIE}
\DeclareAbbreviation\aip{AIP Conf.\ Proc.}
\DeclareAbbreviation\asp{ASP Conf.\ Ser.}

\begin{document}

\chapter{Milky Way Disk}\label{chap1}

\author[1,2]{Daisuke Kawata}%
\author[3]{Robert J.J. Grand}%
\author[4]{Jason A.S. Hunt}%
\author[5]{Ioana Ciuc\u{a}}%

\address[1]{\orgname{Mullard Space Science Laboratory}, \orgdiv{University College London}, \orgaddress{Holmbury St. Mary, Dorking, Surrey RH5 6NT, UK}}
\address[2]{\orgname{National Astronomical Observatory of Japan}, \orgaddress{2-21-1 Osawa, Mitaka, Tokyo 181-8588, Japan}}
\address[3]{\orgname{Astrophysics Research Institute}, \orgdiv{Liverpool John Moores University}, \orgaddress{146 Brownlow Hill, Liverpool L3 5RF, UK}}
\address[4]{\orgname{School of Mathematics \& Physics}, \orgdiv{University of Surrey}, \orgaddress{Guildford GU2 7XH, UK}}
\address[5]{\orgname{Kavli Institute for Particle Astrophysics \& Cosmology (KIPAC)}, \orgdiv{Stanford University}, \orgaddress{Stanford, CA 94305, USA}}

\articletag{Chapter Article tagline: update of previous edition,, reprint..}

\maketitle

\begin{glossary}[Glossary]
\term{$\alpha$-elements} Elements, such as O, Mg, S, Si, Ca, and Ti, whose atomic mass number is a multiple of the mass number of $^4$He, whose nucleus is called $\alpha$-particle. 

\term{[$\alpha$/Fe]} Abundance ratio between $\alpha$-element and iron with respect to that for the Sun in log scale. 

\term{Apo-center} The largest galactocentric radius in an orbit of the star.

\term{Astrophysical stellar parameters} Stellar effective temperature, T$_\mathrm{eff}$, gravity, $\log g$, metallicity, [M/H], and abundances of various elements.

\term{Asymmetric drift} The
difference between the circular speed and the mean rotation velocity of the stars.

\term{Co-rotation Resonance} The condition where the orbital rotation frequency of the star matches that of the non-axisymmetric structure, such as a bar.

\term{Galactic bar} The bar-like stellar over-density structure in the inner galactic disk.

\term{Chemically-defined thick and thin disks} The old high-[$\alpha$/Fe] thick stellar disk and the young low-[$\alpha$/Fe] thin stellar disk.

\term{Circular orbit} The orbit of an object along a complete circle, no radial motion, in the galactic disk. 

\term{Circular speed} The rotation speed of the object in a circular orbit. 

\term{Cosmological simulation} Computer numerical simulation, self-consistently following the formation of the large-scale structure and the formation of the galaxies under an assumed cosmology.

\term{Dynamically self-consistent} Kinematics of stars is consistent with what is expected from the gravitational potential due to their own mass distribution. 

\term{Epicycle frequency} Frequency of the radial oscillation of the stellar orbit. 

\term{Galactoseismology} Studying the kinematic sub-structures of the Milky Way stellar disk and understanding the nature of the Milky Way disk from their dynamical response to perturbations. 

\term{Geometrically-defined thick and thin disk} Geometrically thick and thin stellar disk components. 

\term{Local Standard of Rest (LSR)} A point rotating with a circular speed at the Galactocentric radius of the Sun. 

\term{Moving group} A group of stars sharing a similar position and motion in the phase space. The classical ones are the groups of stars identified in the radial and rotation velocity distribution of the solar neighborhood stars.

\term{Pattern speed} Angular speed of a non-axisymmetric feature rotating in the galactic disk.

\term{Peri-center} The smallest galactocentric radius in an orbit of the star.

\term{Phase space} Six dimensional space of the position and velocity of stars. 

\term{Phase spiral} Spiral arm feature observed in a position and velocity map, such as the vertical  position and velocity map, of the stars.

\term{Proper motion} The angular change in the position of the celestial object in the sky over the time. 

\term{Proper motion of the Sun} The offset velocity of the Sun from the LSR at the Galactocentric radius of the Sun.

\term{Radial migration} The change in the angular momentum of the disk stars.

\term{Orbital Resonance} The condition where orbital epicycle frequency is a simple multiple of the difference in the rotation frequencies of the star and that of the non-axisymmetric structure, such as a bar.

\term{Rigid-body rotation} A feature in the galactic disk rotating at a same angular speed at different radii, so that the shape of the feature does not change in time.

\term{Rotation curve} The circular speed as a function of the galactocentric radius. 

\term{Spiral arms} The spiral-like structure of the over-density of stars and gas in the galactic disk. 

\term{Stellar astrometry} Measurements of the position, including parallax, and motion of the stars.

\term{Stellar parallax} The apparent angle difference in the stellar position when observed from the Earth and from the Sun, which provides the distance of the star by trigonometry.

\end{glossary}

\begin{glossary}[Nomenclature]
\begin{tabular}{@{}lp{34pc}@{}}
BPX-bulge & Boxy/Peanut/X-shape-bulge \\
GSE & Gaia-Sausage-Enceladus \\
GGS & Great Galactic Starburst \\
ILR & Inner Lindblad Resonance \\
LSR & Local Standard of Rest \\
OLR & Outer Lindblad Resonance\\
SNe~II & Type~II supernovae \\
SNe~Ia & Type-Ia supernovae \\
\end{tabular}
\end{glossary}

\begin{abstract}[Abstract]
Our understanding of the Milky Way disk is rapidly improving with the recent advent of the high quality and vast amount of observational data. We summarize our current view of the structure of the Milky Way disk, such as the masses and sizes of the gas and stellar disks, and the position and motion of the Sun in the disk. We also discuss the different definitions of the thick and thin disks of the Milky Way, the non-axisymmetric structures of the stellar disk, such as the bar and spiral arms, and the radial migration which can be triggered by these non-axisymmetric stellar structures. After the revolutionary data from the European Space Agency's Gaia mission, our view of the Milky Way disk has been transformed to a non-equilibrium system with many complicated structures in stellar kinematic distribution. We also summarize the recent findings of Galactoseismology research. These detailed observational data provide the archaeological information for us to unveil the formation and evolution history of the Milky Way disk, with the aid of the high-resolution numerical simulations of the Milky Way-like galaxy formation. We also discuss the current view of the formation history of the Milky Way disk.
\end{abstract}

\begin{BoxTypeA}[chap1:box1]{Key Points}
\begin{itemize}
\item The Milky Way is a barred spiral disk galaxy. The Milky Way disk has a stellar bar in the inner disk, and also has spiral arms. The number and shape of the spiral arms are still unknown.
\item The Milky Way disk consists of the gas and stellar disks. The stellar disk shows the geometrically thick and thin disks. Also, there are two distinct stellar populations with different chemical abundances, high and low [$\alpha$/Fe] abundances, corresponding to thick and thin disks, respectively. These geometrically-defined thick and thin disks are not same as the chemically-defined thick and thin disks.
\item The Milky Way disk is not completely in dynamically equilibrium. There are many kinematic sub-structures, such as ridges and corrugation features, observed. The outer disk shows the upward and down-ward warps in the opposite sides of the disk. 

\item The main source of the perturbation is considered to be the Sagittarius dwarf galaxy, which is falling into the Milky Way. However, the exact mechanism causing the observed stellar kinematic features are still in debate. 

\item The Milky Way disk formation likely started with gas-rich mergers at the early epoch of the Universe, which built the thick disk. Then,  following a long term quiet phase of no significant merger helped to build the thin disk. The last most significant merger, the Gaia-Sausage-Enceladus merger, likely impacted the transition from the thick disk to the thin disk formation phase. 
\end{itemize}
\end{BoxTypeA}

\section{Introduction}\label{chap1:Intro}

One of the most striking and fascinating features visible in the night sky is the stream of stars overarching the sky. This is merely an edge-on view of the stellar disk of our Galaxy, and from this view the name of the Milky Way derives. Thus, understanding the structure of the Milky Way disk and its formation history has been a fundamental topic of astronomy since ancient times. 

Our understanding of the Galactic disk has been dramatically enhanced due to the advent of ground-based and space-based observational facilities covering all the wavelengths of light from the gas and stars, and powerful computing facilities. In particular, the European Space Agency (ESA)'s Gaia mission \citep{Gaia+Prusti16} has revolutionized our view of the Milky Way disk. The Gaia mission launched in 2013 is the successor to ESA's Hipparcos mission (1989$-$1983) which was the first space astrometry mission, and measured parallaxes and proper motions for approximately 118,000 bright stars. The sharp image of stars obtained in space allows space missions to precisely measure the stars' positions in the sky. The positions of these stars in the sky vary with the Earth's orbit around the Sun, and the magnitude of these differences for a given star, otherwise known as the trigonometric parallax angle, gives the distance to the star. In addition, the long time baseline of the observations allows the measurement of the movement of stars, i.e. the proper motion of the stars. Although the final data release of the Gaia mission is planned in the 2030s, the Gaia mission has made several intermediate data releases already. Their 3rd data release (Gaia~DR3) provides the precise positions, parallaxes, and proper motions for about 1.5 billion stars up to $G\sim21$~mag with about 100 times better accuracy than what was achieved by Hipparcos. In addition, the radial velocities measured with the onboard spectrograph, RVS, provide radial velocities for about 33 million stars brighter than $G_\mathrm{RVS}\sim14$~mag. Astrophysical parameters of stars, such as effective temperature, T$_\mathrm{eff}$, gravity, $\log g$, metallicity, [M/H], and abundances of various elements are also measured with the RVS spectra and the onboard BP/RP spectro-photometric data. 

Spectroscopic stellar surveys, such as the RAdial Velocity Experiment (RAVE), the Sloan Digital Sky Survey (SDSS), the Sloan Extension for Galactic Understanding and Exploration (SEGUE), the Apache Point Observatory Galactic Evolution Experiment (APOGEE), the Large Sky Area Multi-Object Fiber Spectroscopic Telescope (LAMOST), the Gaia-ESO survey, the Galactic Archaeology with HERMES survey (GALAH), and Hectochelle in the Halo at High Resolution survey (H3) are adding value to the Gaia data by providing further, more detailed chemical abundance information for a portion of the Gaia stars. In addition, the cold gas, including dust and molecular gas, neutral hydrogen gas, and the warm and hot gas distribution in the Milky Way are revealed by the X-ray to radio observations. These observational data are compared with theoretical models and our understanding of the current structure and formation history of the Milky Way disk has significantly improved recently. 

This chapter summarizes our current understanding of the Milky Way disk. Section~\ref{sec:overall-structure} describes the overall structure of the Milky Way disk unveiled by the Gaia mission. Section~\ref{sec:formation} shows the current view of the formation process of the Milky Way disk. Section~\ref{sec:summary} provides the summary of the chapter and future prospects. 

\begin{figure}[t]
\centering
\includegraphics[width=\textwidth]{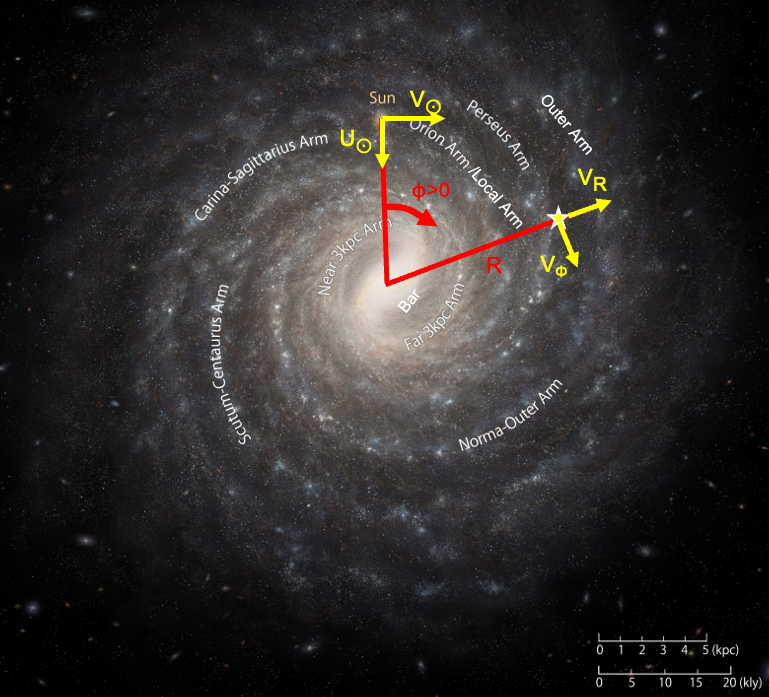}
\caption{Artist impression of the Milky Way disk in the face-on view. In this view, the Milky Way disk is rotating clockwise. The arrows from the Sun indicate the directions of the Sun's radial, $U_{\odot}$, and azimuthal, $V_{\odot}$, proper motion. The yellow star symbol presents an example star. The yellow arrows from the star show the directions of their radial, $V_\mathrm{R}$, and azimuthal, $V_\mathrm{\phi}$, velocities. The red line from the center of the Milky Way to the star indicates the Galactocentric radius of the star, $R$. The red line from the center to the Sun is where the azimuthal angle, $\phi=0$. The red arrow shows the direction of the positive $\phi$. The bar and the name of the spiral arms are also highlighted.  Modified from NAOJ, \citet{Nakanishi+Sofue16}, \citet{Reid+Menten+Brunthaler+19}, and \citet{Baba+Kawata+Schoenrich22}.}
\label{fig:mwd-image}
\end{figure}

\section{The structure of the Milky Way Disk}
\label{sec:overall-structure}

Figure~\ref{fig:mwd-image} shows an artist's impression of the overall structure of the Milky Way disk as viewed face-on. The Milky Way is believed to be a barred spiral disk galaxy. It is still challenging to capture the true structure of the whole Milky Way disk. This is because we live within the Milky Way disk, and it is impossible to see the whole Milky Way from outside. Although we can measure the position of individual stars in the Milky Way disk, accurately measuring the distance to the faint and/or distant stars is always a challenge in astronomy. 
In addition, the interstellar dust with complex structure blocks starlight in complicated ways. Therefore, it is not straightforward to obtain the true density distribution of the stars in the Milky Way disk. Still, our understanding of the structure of the Milky Way disk is rapidly improving with data from Gaia and other complementary Milky Way surveys. 

In this section, we provide an overall 
picture of the Milky Way disk in our current view. The basic structure of the Milky Way is also summarized in \citet{Bland-Hawthorn+Gerhard16} and \citet{McMillan17}. A thorough review of the stellar kinematics of the Milky Way is provided in \citet{Hunt+Vasiliev24}. A comprehensive Milky Way stellar distribution model which describes all the disk, bulge and halo stars was constructed by \citet{Robin+Reyle+Derriere+03} and is known as the Besan\c{c}on Galaxy Model\footnote{\url{https://model.obs-besancon.fr/}}. The Besan\c{c}on Galaxy Model is regularly updated and the latest model \citep{Robin+Bienayme+Salomon+22}, which includes the dark matter halo, is adjusted to match the Gaia data and is dynamically self-consistent: the kinematics of stars are consistent with what is expected from the gravitational potential due to their mass distribution. The model provides the mass and size of 8 different disks with different masses and ages: the oldest one corresponds to the thick disk (Section~\ref{sec:thin-thick}). The Besan\c{c}on Galaxy Model is one of the valuable reference models of the whole Milky Way.

Section~\ref{sec:ms-mwd} summarizes the mass and size of the different components of the Milky Way disk. In Section~\ref{sec:thin-thick}, we describe the thick and thin disk stellar populations. Sections~\ref{sec:bar} and \ref{sec:spirals} describe the bar and spiral structures, respectively. Section~\ref{sec:radial-migration} describes the radial migration process induced by the bar and spiral arm structures. Section~\ref{sec:galseismo} provides the current status of Galactoseismology, which studies how the Milky Way responds to past dynamical impacts through observed kinematic sub-structures in the Milky Way disk stars.

\subsection{Mass and size of the Milky Way disk}
\label{sec:ms-mwd}


\subsubsection{Stellar disk}

The overall stellar mass of the Milky Way disk is estimated to be about $M_\mathrm{\star,disk}=3-5\times10^{10}$~M$_{\odot}$ \citep[e.g.,][]{Licquia+Newman16,Binney+Vasiliev24}. The surface mass density profile of the galactic disk is generally well described with an exponential profile, 
\begin{equation}
    \Sigma(R)=A \exp(-R/R_\mathrm{d,\star}),
\label{eq:exp-profile}
\end{equation}
where $R_\mathrm{d,\star}$ is the scale length of the disk and $R$ is the Galactocentric radius in the cylindrical coordinate. Compiling the data from the literature, \citet{Licquia+Newman16} estimate $R_\mathrm{d,\star}\sim2.6$~kpc. 
However, the scale length of the Milky Way disk as defined in this way is much smaller compared to other disk galaxies of similar stellar mass to the Milky Way. \citet{Lian+Zasowski+Chen+24} argue that the radial density profile of the Milky Way depends on stellar age, and that they are not well-described with a single exponential profile, but are rather better explained with broken-exponential profiles. They obtain the half-light radius (the radius which contains half of the total disk starlight) to be $R_{50}=5.75\pm0.38$~kpc, which is consistent with other external disk galaxies of similar mass to the Milky Way. 

\subsubsection{Gas disk}

The Milky Way disk also contains a gas disk. Compiling the literature data, \citet{McMillan17} estimates the mass and size of the Milky Way gas disk. The author uses the following function to describe the mass distribution of H\,\textsc{i} gas and molecular gas disks in cylindrical coordinates.
\begin{equation}
    \rho_\mathrm{g}(R,z)=\frac{\Sigma_\mathrm{g}}{4 z_\mathrm{d,g}}\exp\left(-\frac{R_\mathrm{m}}{R}-\frac{R}{R_\mathrm{d,g}}\right)\mathrm{sech}^2(z/2z_\mathrm{d,g}).
\label{eq:gasd-profile}
\end{equation}
This is similar to the exponential profile (equation~\ref{eq:exp-profile}) with scale length $R_\mathrm{d,g}$ in the outer disk, but the density declines with decreasing $R$ in the inner disk with a peak at $R=\sqrt{R_\mathrm{m}R_\mathrm{d,g}}$. The vertical density profile is described with sech$^2$ function with the scale height of $z_\mathrm{d,g}$. \citet{McMillan17} suggests that the mass of HI disk is $M_\mathrm{H\,\textsc{i}}=1.1\times10^{10}$~M$_{\odot}$ with the parameters of $R_\mathrm{d,H\,\textsc{i}}=7$~kpc, $R_\mathrm{m,H\,\textsc{i}}=4$~kpc, and $z_\mathrm{d,H\,\textsc{i}}=0.085$~kpc, and the mass of molecular gas disk is $M_\mathrm{H\,2}=1.2\times10^{10}$~M$_{\odot}$ with the parameters of $R_\mathrm{d,H\,2}=1.5$~kpc, $R_\mathrm{m,H\,2}=12$~kpc, and $z_\mathrm{d,H\,2}=0.045$~kpc.

\subsubsection{Sun's location and motion in the Milky Way disk}

The Sun is located near the edge of the Milky Way disk, as shown in Figure~\ref{fig:mwd-image}. The distance to the Sun from the Galactic center, $R_\mathrm{0}=8.275\pm0.09_\mathrm{stat.}\pm0.33_\mathrm{sys.}$~kpc, is accurately measured using 16 years of the orbital data of the star S2 around the supermassive black hole, Sagittarius~A$^*$ (Sgr~A$^*$), at the center of the Milky Way by \citet{Gravity+Abuter+Amorim+21}. Here, $_\mathrm{sys.}$ and $_\mathrm{stat.}$ indicate systematic and statistical uncertainties, respectively. The systematic uncertainty is more difficult to assess than the statistical uncertainty. Unless otherwise stated, the uncertainties shown in this chapter indicate the latter. 

\citet{Reid+Brunthaler20} measure the proper motion of the radio emission from Sgr~A$^*$ with respect to two extra-galactic (therefore stationary) radio sources using 18 years of the Very Long Baseline Array data. They find negligible intrinsic motion of Sgr~A$^*$, i.e. the proper motion of Sgr~A$^*$ is reflecting the motion of the Sun in the Milky Way. 

The Sun is rotating around the Milky Way disk at the Galactocentric radius of $R_\mathrm{0}$. The Sun's proper motion is defined as the offset of Sun's velocity from the Local Standard of Rest (LSR). The LSR is defined as a point rotating with a circular speed, $\Theta_0$, at $R_\mathrm{0}$, i.e. having a completely circular orbit without any radial motion. The radial, azimuthal, and vertical components of Solar motion with respect to LSR are denoted with $U_{\odot}$, $V_{\odot}$, and $W_{\odot}$, respectively (Figure~\ref{fig:mwd-image}). The above-mentioned proper motion of Sgr~A$^*$ provides the Sun's total motion in the Galactic rest frame. Hence, the longitudinal proper motion of Sgr~A$^*$ corresponds to the total rotation speed of the Sun, $\Theta_0+V_{\odot}$. Using this and $R_0$ measured by \citet{Gravity+Abuter+Amorim+21} as the strong priors, \citet{Almannaei+Kawata+Baba+24} fit the kinematics of the young stars in the Gaia Data Release 3 \citep[Gaia~DR3,][]{GaiaDR3+Vallenari+Brown+23} with an axisymmetric disk model. The young stars, such as O-type and B-type stars (OB stars), are expected to closely follow the circular motion. Hence, their asymmetric drift, the difference between the mean rotation velocity of the stars and the circular velocity, is very small, and they are an ideal tracer to measure the circular speed. Using the Gaia~DR3 OB stars' data, \citet{Almannaei+Kawata+Baba+24} obtain $\Theta_0=234\pm2$~km~s$^{-1}$, $\Theta_0+V_{\odot}=248\pm2$~km~s$^{-1}$, and $U_{\odot}=-9.65\pm0.43$~km~s$^{-1}$, which are consistent with the other studies. Comparing with the numerical simulations of a disk galaxy model with a bar and spiral arms, \citet{Almannaei+Kawata+Baba+24} note that the stellar kinematics are affected by the non-axisymmetric structures, such as the bar and the spiral arms (see more information in Sections~\ref{sec:bar} and \ref{sec:spirals}), and therefore it is challenging to accurately measure the circular speed of the Milky Way. However, they demonstrate that taking the stellar kinematics in a larger region, e.g. within 2~kpc of the Sun, mitigates the issue. 

The proper motion of Sgr~A$^*$ in the latitudinal direction reflects the Sun's motion toward the North Galactic pole, which is transformed to $W_{\odot}=8.57\pm0.27$~km~s$^{-1}$. The Sun is located above the mid-plane of the Milky Way disk. From Gaia~DR2 data, \citet{Bennett+Bovy19} measure the distance of the Sun from the Galactic mid-plane to be $z_0=20.8\pm0.3$~pc.

\subsubsection{Local dark matter density from the Milky Way disk kinematics}
\label{sec:ldmd}

Another important component contributing to the gravitational potential in the Milky Way disk is dark matter. The stellar and gas kinematics follows the total gravitational potential of the star, gas, and dark matter. Analyses of the total density of the local volume and finding the missing mass density, comparing it with the visible mass density, date back to 1920s, such as \citet{Kapteyn22} and \citet{Jeans22}. The reviews of the various methods using the modern data are found in \citet{Read14} and \citet{deSalas+Widmark21}. \citet{deSalas+Widmark21} compiled the post-Gaia measurements of the local dark matter density and highlighted that the various measurements of the local dark matter density at the Solar position show a wide range of values $\rho_\mathrm{DM,\odot}=0.008-0.016$~M$_{\odot}$~pc$^{-3}$. 
The reason for the discrepancy between the methodologies is likely due to the required assumption of equilibrium of the disk kinematics. As discussed in Section~\ref{sec:galseismo}, the Gaia data have uncovered the disequilibrium of the Milky Way disk, which inevitably impacts the accuracy of the measurements of the gravitational potential from the disk kinematics. 

\subsubsection{Rotation curve}
\label{sec:rc}

The kinematics of the stars and gas of the Milky Way disk can provide the circular speed as a function of the Galactocentric radius, the so-called the rotation curve \citep{Sofue20}. The mean rotation velocity of stars, $<v_{\phi}(R)>$, is usually slower than the circular speed, $\Theta(R)$, and this difference is called asymmetric drift, as mentioned above. The amount of asymmetric drift depends on the radial velocity dispersion profile, $\sigma_\mathrm{R}(R)$, and the density profile of the tracer stars, $\Sigma(R)$.  
Hence, to measure the rotation curve, we need to accurately measure both the density profile and kinematics of the tracer stars. Recent studies of measuring the Milky Way rotation curve based on the Gaia data and ground-based spectroscopic survey data reach up to about $R=30$~kpc. Interestingly, several studies \citep[e.g.,][]{Wang+Chrobakova+Lopez-Corredoira+23} find a declining rotation curve at $R>\sim17$~kpc. This indicates lower dark matter contribution than what is expected from the standard dark matter density profile for the Milky Way halo whose mass is around $M_\mathrm{tot}=10^{12}$~M$_{\odot}$ and inferred with the larger radii tracers, such as globular clusters and satellite galaxies \citep[e.g.,][for a review]{Wang+Han+Cautun+20,Hunt+Vasiliev24}. This tension remains a puzzling issue. However, we note that accurately obtaining the density profile, especially in the outer disk, is challenging, and studies use the extrapolation of an exponential profile, like equation~(\ref{eq:exp-profile}), with the scale length and scale height inferred in the inner disk. Hence, more observational data in the outer disk would be crucial to resolve this tension.

\subsection{Thick and Thin disks}
\label{sec:thin-thick}

From the vertical number density profile of stars, \citet{Yoshii82} and \citet{Gilmore+Reid83} found a separate, thicker disk component to the main stellar thin disk. This extra component is called the thick disk. Each of these stellar disk density distributions is described with an exponential profile in both radial and vertical directions as the following equation.
\begin{equation}
    \Sigma_\mathrm{\star}(R,z)=\frac{\Sigma_\mathrm{0,\star
    }}{2 z_\mathrm{d,\star}}\exp\left(-\frac{R}{R_\mathrm{d,\star}}-\frac{|z|}{z_\mathrm{d,\star}}\right),
\label{eq:sd-rvexp}
\end{equation}
where $R_\mathrm{d,\star}$ is called the scale length and $z_\mathrm{d,\star}$ is called the scale height. Please note that the scale height in the exponential law is different from the scale height in $\mathrm{sech}^2$ profile used for the gas density profile in equation~(\ref{eq:gasd-profile}). In this description, the total mass of the component can be conveniently described as $M_\mathrm{\star}=2\pi\Sigma_\mathrm{0,\star}R_\mathrm{d,\star}^2$. Analyzing the number counts of the stars using the photometric data of SDSS, \citet{Juric+Ivezic+Brooks+08} obtain $R_\mathrm{d,thick}=3.6$~kpc and $z_\mathrm{d,thick}=0.9$~kpc for the thick disk, while $R_\mathrm{d,thin}=2.6$~kpc and $z_\mathrm{d,thin}=0.3$~kpc for the thin disk. Following \citet{Kawata+Chiappini16}, we call this definition of thick and thin disks "geometrically-defined" thick and thin disks. 

On the other hand, the thick and thin disk stellar populations are often defined by their distinct chemical compositions. The most fundamental way of characterizing the chemical abundances of the stars is using the iron abundance with respect to the hydrogen abundance, [Fe/H], and the $\alpha$-element abundances with respect to the iron abundance, [$\alpha$/Fe]. Here, [X/Y] means that the ratio of X and Y element abundance in log scale with respect to their abundance in the Sun, i.e.
\begin{equation}    
\mathrm{[X/Y]}=\log(\mathrm{Z_X/Z_Y})_{\star}-\log(\mathrm{Z_X/Z_Y})_{\odot}.
\end{equation}
Here, $\mathrm{Z_X}$ means the mass fraction of X~element with respect to the star's total mass. For example, for the Sun, the hydrogen abundance is $Z_{H,\odot}=0.7438$ \citep{Asplund+Amarsi+Grevesse21}. $\alpha$-elements include O, Mg, S, Si, Ca, and Ti, whose atomic mass number is a multiple of the mass number of $^4$He, whose nucleus is called $\alpha$-particle. The definition of $\alpha$-element abundance often depends on the surveys, because different surveys use different spectrographs with different wavelength ranges and adopt different absorption lines and/or spectrum fitting methods to measure the abundance. For example in APOGEE DR17 \citep{Majewski+Schiavon+Frinchaboy+17}, $\alpha$-abundance is a combination of the abundance of O, Mg, S, Si, Ca, and Ti, and the contribution from these different elements depends on the stellar parameters, such as effective temperature, $T_\mathrm{eff}$\footnote{\url{https://www.sdss4.org/dr17/irspec/parameters/}}. Nevertheless, [$\alpha$/Fe] is a good indicator of when stars were born. This is because $\alpha$-elements are predominantly produced by Type~II supernovae (SNe~II) from massive ($m_{\star}>\sim10$~M$_{\odot}$) stars whose lifetime is less than 10~Myr. On the other hand, Fe are more significantly produced by Type~Ia SNe (SNe~Ia), which may be from a merger of low-mass binary stars, whose lifetimes are in a wide range between 0.1 and 20 Gyr \citep[e.g.,][]{Kobayashi+Karakas+Lugaro20}. However, the exact progenitors of SNe~Ia are still not well known. Consequently, stars are high-[$\alpha$/Fe] abundance if they are born at a time when SNe~II are dominant, i.e. during an active star formation phase. On the other hand, low-[$\alpha$/Fe] stars can form when the star formation rate is decreasing or is at a lower level for a long time, because SNe~Ia become more dominant than SNe~II \citep[e.g.,][]{Mason+Crain+Schiavon+24}. 

\begin{figure}[t]
\centering
\includegraphics[width=0.75\textwidth]{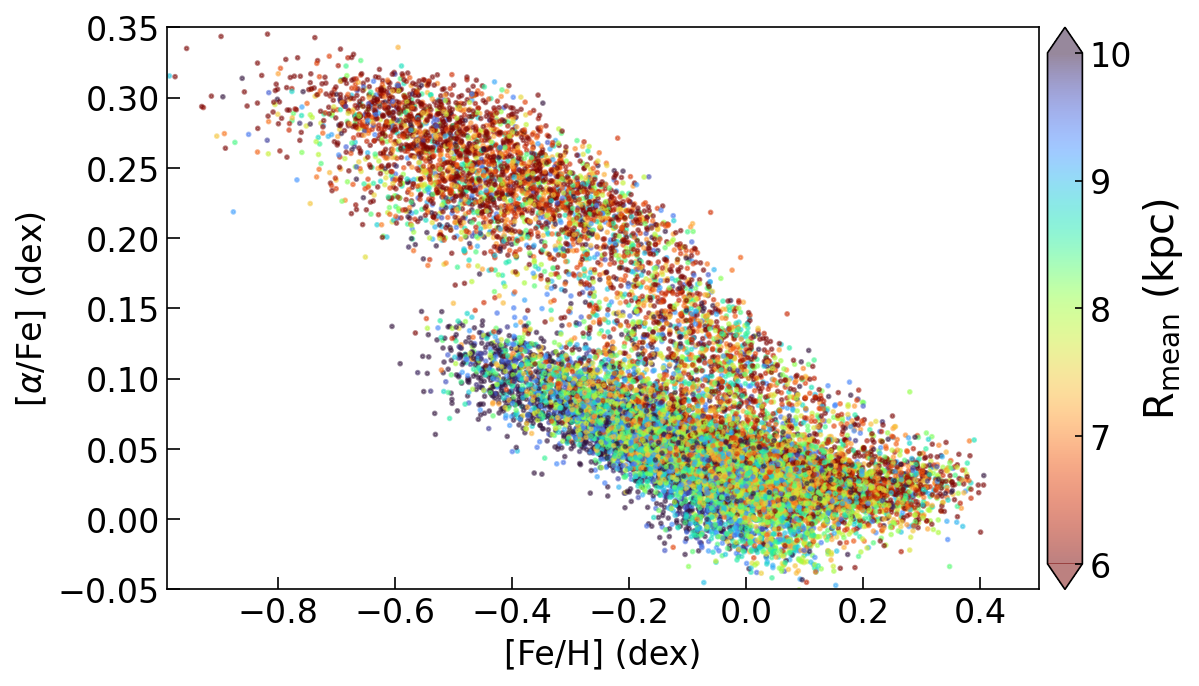}
\caption{[Fe/H] and [$\alpha$/Fe] distribution of APOGEE giant stars in $R<15$~kpc and $|z|<3$~kpc color-coded with the mean orbital radius, $R_\mathrm{mean}$. }
\label{fig:feh_alphafe_rcol}
\end{figure}

Figure~\ref{fig:feh_alphafe_rcol} shows [Fe/H] and [$\alpha$/Fe]\footnote{APOGEE~DR17 provides [$\alpha$/M] rather than [$\alpha$/Fe], and "M" indicates the total metal abundance. However, to avoid a confusion, we call [$\alpha$/M] in APOGEE data [$\alpha$/Fe], since [Fe/H] is very similar to [M/H].} distribution of APOGEE~DR17 giant stars from \citet{Ciuca+Kawata+Ting+24}'s high-confidence age sample.
As seen in Figure~\ref{fig:feh_alphafe_rcol}, high-resolution spectroscopic surveys of the Milky Way disk stars clearly show that there are two chemically distinct populations of stars, high-[$\alpha$/Fe] and low-[$\alpha$/Fe] populations in the large region of the Milky Way disk \citep[e.g.,][]{Hayden+Bovy+Holtzman+15,Imig+Price+Holtzman+23}. The high-[$\alpha$/Fe] stars tend to have significantly higher velocity dispersion than the low-[$\alpha$/Fe] stars \citep[e.g.,][]{Sun+Huang+Shen+24}. Hence, the high-[$\alpha$/Fe] and low-[$\alpha$/Fe] stars are also called thick and thin disks, respectively. Following \citet{Kawata+Chiappini16}, we call these "chemically-defined" thick and thin disks. However, the chemically-defined thick and thin disks are not completely the same as the geometrically-defined thick and thin disks. For example, \citet{Bensby+Feltzing+Oey14} show that the stars in the geometrically-defined thick disk are mainly in high-[$\alpha$/Fe] population, but they are also spread in low-[$\alpha$/Fe] population. \citet{Bovy+Rix+Liu+12} show that the high-[$\alpha$/Fe] disk has a smaller scale length than the low-[$\alpha$/Fe] disk. Figure~\ref{fig:feh_alphafe_rcol} also shows that the high-[$\alpha$/Fe] disk stars in the local disk tend to come from the inner disk, whose mean orbital radius, $R_\mathrm{mean}$, is smaller. Here, $R_\mathrm{mean}$ is computed from the orbital integration under the Milky Way potential and defined as $0.5(R_\mathrm{apo}+R_\mathrm{peri})$, where $R_\mathrm{apo}$ and $R_\mathrm{peri}$ are the apo-center, the largest radius reached by the orbit of the star, and the peri-center, the smallest radius of the orbit. The orbits are computed with a galactic dynamics software package, \texttt{galpy}\footnote{\url{https://www.galpy.org/}} \citep{Bovy15}, and the distance to the stars are taken from \citet{Queiroz+Anders+Chiappini+23}. This is contradictory to the larger scale length of the geometrically-defined thick disk than the thin disk as mentioned above. Analyzing the numerical cosmological simulation result of a Milky Way-sized disk galaxy, \citet{Rahimi+Carrell+Kawata14} find that this can be explained by the natural consequence of the cosmological formation history of galaxies. As we explain more in Section~\ref{sec:formation}, the high-[$\alpha$/Fe] disk forms at an early epoch as a thick and compact disk, while the low-[$\alpha$/Fe] disk forms in the later phase as a thin and larger disk, but flaring (larger scale-height) at the outer radii. Hence, the low-[$\alpha$/Fe] stars contribute to the geometrically thick disk, and make the geometrically thick disk larger. In fact, \citet{Bensby+Feltzing+Oey14} demonstrate that the chemically-defined thick and thin disks are more correlated with the age, and the chemically-defined thick disk is generally older than the chemically-defined thin disk. They also find that the high-[$\alpha$/Fe] stars are more centrally concentrated than low-[$\alpha$/Fe] stars. Although stellar ages are difficult to measure, \citet{Bensby+Feltzing+Oey14} obtain the age for the turn-off stars and sub-giants using the stellar parameters, $\log g$, $T_\mathrm{eff}$ and [Fe/H], precisely measured from high-resolution high signal-to-noise spectroscopic data. This age trend of the old chemically-defined thick and younger thin disks is also confirmed by the highly precise asteroseismology ages \citep[e.g.,][]{Miglio+Chiappini+Mackereth+21}. Hence, it is better to define the two populations of the disks with age to associate with their formation history. However, since it is difficult to obtain the precise age for many stars, it is more convenient to define the distinct stellar populations of thick and thin disks using their chemical properties. Hereafter in this chapter, when we discuss the thick and thin disks, we indicate the chemically-defined old high-[$\alpha$/Fe] thick disk and young low-[$\alpha$/Fe] thin disk, respectively.


\subsection{Bar and Spiral Arms}
\label{sec:bar-spirals}

\subsubsection{Bar}
\label{sec:bar}

As shown in Figure~\ref{fig:mwd-image}, the Milky Way disk harbors a prominent bar structure in the inner disk region. However, the shape, angle, length, pattern speed, and formation time of the bar are all still in debate. The bar of the Milky Way was first suggested to explain the anomalous motion of the gas in the inner disk \citep{Peters75}. The bar structure is also inferred from the asymmetry of the number of infrared bright stars \citep{Nakada+Onaka+Yamamura+91}, and the near-infrared (NIR) emission distribution of the inner disk stars \citep{Blitz+Spergel91b}.

\citet{Blitz+Spergel91b} note that the Cosmic Microwave Background Explorer (COBE) NIR image clearly shows the ``peanut-shaped" bulge in the central region of the Milky Way disk. 
The red clump stars are in the core-helium burning phase of metal-rich stars with a specific absolute luminosity, and hence the red clump stars are often used as the standard candle to map the stellar structure of the Galactic disk. \citet{McWilliam+Zoccali10} find two groups of red clump stars at different distances in the inner bulge region, interpreting it as the X-shape of the inner bulge. Utilizing the large NIR photometric survey of the Galactic disk stars, VISTA Variables in the Via Lactea (VVV) survey, \citet{Wegg+Gerhard13} map the red clump stars' density distribution, and reconstruct the clear Boxy/Peanut/X-shape (BPX) bulge structure. The major axis of the BPX-bulge extends up to about $R_\mathrm{BPX}=2$ kpc with the exponential scale length of each axes being (0.70, 0.44, 0.18) kpc along $x_\mathrm{BPX}$, $y_\mathrm{BPX}$, and $z_\mathrm{BPX}$, where $x_\mathrm{BPX}$ is the major axis, and $x_\mathrm{BPX}-y_\mathrm{BPX}$ is in the disk plane. Because of the Peanut/X-shape, the vertical scale-height increases to 0.46~kpc at $x_\mathrm{BPX}=1.725$~kpc. 

Although the connection between the BPX-bulge and the outer long bar was debated, analyzing the VVV data and using an N-body model as a prior, \citet{Wegg+Gerhard+Portail15} convincingly show that the BPX-bulge is the inner region of the long bar, and the long bar extends to the radius of $R_\mathrm{lb}=5.0\pm0.2$~kpc. They also measure the angle of the bar from the line of the Sun and the Galactic center, and obtain the angle of $\phi_\mathrm{lb}=28^{\circ}-33^{\circ}$. The long bar is also visible in the Gaia data combined with the APOGEE survey data \citep{Queiroz+Chiappini+Perez-Villegas+21}. Also, \citet{Zhang+Belokurov+Evans+24} show that the bar is visible with the Gaia data's low-amplitude long variable stars. They suggest that the bar length is about $R_\mathrm{lb}=4$~kpc, and obtain the bar angle of $\phi_\mathrm{lb}=25^{\circ}$ which is similar to \citet{Wegg+Gerhard+Portail15}. However, the end of the bar could overlap with the spiral arms, which affects the apparent size of the bar and the angle of the bar. Hence, the measurement and/or definition of the length of the bar and the angle of the bar are not straightforward to measure \citep[e.g.,][]{Vislosky+Minchev+Khoperskov+24}, and are still in debate. 

\paragraph{Pattern speed of the bar}

The stars in the bar region in the inner disk show a cylindrical rotation \citep[e.g.][]{Howard+Rich+Reitzel+08}. This also means that the bar is a dominant structure in the inner disk, and the bar has a rigid-body rotation to keep its shape. Based on a comparison with a numerical simulation, \citet{Shen+Rich+Kormendy+10} estimate that the mass of the Milky Way bulge, i.e. nearly-spherical non-rotating stellar component, should be less than 8~\% of the disk mass. 

The pattern speed of the bar can be measured in several ways. The first convincing measurement was made by modelling the kinematics of the stars in the solar neighborhood. Several moving groups are identified in the radial and rotation velocity distribution of the stars in the solar neighborhood (Section~\ref{sec:galseismo}). The Hercules stream is a prominent moving group, which shows slower rotation than the mean rotation speed of the other stars and moves outward in the Galactic disk. If the orbital angular frequency of $\Omega$ (the angular speed of stars in the azimuthal direction) and the epicyclic frequency, $\kappa$, are in relation with the pattern speed of the bar, $\Omega_\mathrm{b}$, as $\Omega_\mathrm{b}=\Omega+\kappa/2$, which is called the Outer Lindblad Resonance (OLR), 
the orbit of the star has a closed ellipse orbit in the rotating frame of the bar. In this resonance orbit, a star completes exactly 2 radial oscillations, i.e. passing pericenter and apocenter twice for every rotation around the bar. \citet{Dehnen99} showed that Hercules stream could be caused by the bar's OLR.  If true, this would indicate the pattern speed of the bar to be about $\Omega_\mathrm{b}=53$~km~s$^{-1}$~kpc$^{-1}$. However, the Hercules stream can also be created by other resonances of the bar, including the co-rotation resonance, i.e. $\Omega_\mathrm{b}=\Omega$, \citep{Perez-Villegas+Portail+Wegg+Gerhard17} and the 4:1 resonance, i.e. $\Omega_\mathrm{b}=\Omega+\kappa/4$, \citep{Hunt+Bovy18}, which would imply a slower pattern speed. Alternatively, Hercules stream could be created (or interfered with) by transient spiral structure (Section~\ref{sec:spirals}), limiting our ability to measure the bar pattern speed from local data \citep[e.g.][]{Hunt+Hong+Bovy18,Hunt+Bub+Bovy19}. 

More recent studies converge to a slower bar pattern speed. The kinematics of the gas in the bar region favors a slower pattern speed of about $\Omega_\mathrm{b}=40$~km~s$^{-1}$~kpc$^{-1}$ \citep{Sormani+Binney+Magorrian15}. Also, thanks to NIR photometric and spectroscopic surveys, the kinematics of the stars in the bar are now directly measured. Comparison with kinematic models also suggests a lower pattern speed of around $33-40$~km~s$^{-1}$~kpc$^{-1}$ \citep[e.g.,][]{Bovy+Leung+Hunt+19,Sanders+Smith+Evans19,Clarke+Gerhard22}. This slow pattern speed corresponds to the case of explaining the Hercules stream with co-rotation resonance of the bar. 

Furthermore, \citet{Chiba+Friske+Schoenrich21} suggest that the kinematics of the Galactic disk stars can be used to measure the slowing-down of the pattern speed of the bar. Slowing down of the pattern speed of the bar is often seen in N-body simulations and it is a natural consequence of the dynamical friction transferring angular momentum from the bar to the dark matter halo. They also suggest that the current pattern speed is around $35$~km~s$^{-1}$~kpc$^{-1}$, but that it was faster in the past. 

\paragraph{Age of the bar}
\label{sec:bar-age}

\citet{Bovy+Leung+Hunt+19} measure the age and stellar abundance of the stars in the Galactic bar from the APOGEE data. They find that the stars in the bar are dominated by the old high-[$\alpha$/Fe] stars like the thick disk. Hence, \citet{Bovy+Leung+Hunt+19} suggest that the bar in the Milky Way formed at the same epoch as the formation of the thick disk. However, the age of the stars in the bar does not equate to the age of the bar, because stars older and younger than the formation age of the bar can be captured by the bar. 

Bar formation is well known to induce gas inflow to the central sub-kpc of the galaxy, and create the nuclear gas disk. The stars formed in the nuclear gas disk build the nuclear stellar disk. Using a hydrodynamical galaxy evolution simulation including gas dynamics and star formation, \citet{Baba+Kawata20} demonstrate that the oldest age of the nuclear disk stars becomes consistent with the age of the bar, because they start forming after the bar forms. 

\citet{Sanders+Kawata+Matsunaga+24} identify the Mira variable stars in the nuclear stellar disk of the Milky Way using the VVV data. Using the age-period relation of Miras calibrated in \citet{Zhang+Sanders23}, they find that the Miras younger than about 8~Gyr show colder kinematics, i.e. their velocity dispersion is smaller. They conclude that the nuclear stellar disk started forming about 8~Gyr ago, and that it is also the time when the bar formed. This is a relatively less ambiguous measurement of the age of the bar, and it indicates the bar formed at an early epoch of the formation of the disk. 

\subsubsection{Spiral arms}
\label{sec:spirals}

Figure~\ref{fig:mwd-image} shows the currently-believed spiral arm structure of the Milky Way. Note that this image is created by the extrapolations of the structures of the spiral arms inferred from a relatively small number ($\sim200$) of high-mass star-forming regions, whose distances are accurately measured by the radio Very Long Baseline Interferometer (VLBI) \citep{Reid+Menten+Brunthaler+19}. The nearby stellar over-density structures of the spiral arms are also seen in the young OB stars in the Gaia data \citep[e.g.,][]{Poggio+Drimmel+Cantat_Gaudin+21}, red clump stars \citep{Lin+Xu+Hou+22}, Cepheid variable, and HII regions \citep[e.g., see][for a review]{Hou21}. Also, the spiral arms are identified with the systematic kinematical features \citep{Eilers+Hogg+Rix+20}. However, the location or pitch angle of these arms are not always consistent with each other. Hence, the number of arms and the location of the arms are still under debate, and it remains challenging to identify spiral arm structure in the Milky Way disk. 

The nature and the origin of the spiral arms remain controversial \citep[e.g., see][for a review]{Dobbs+Baba14}. For a long time, the spiral arms have been believed to be a density-wave, where the stellar density-wave features are in a rigid-body rotation, despite the differential rotation of the stars \citep{Lin+Shu64}. \citet{Kalnajs73} suggested a similar scenario called kinetic density-wave theory, which uses the closed ellipse orbit in the rotating frame with $\Omega_\mathrm{p}=\Omega-\kappa/2$, which is called the Inner Lindblad resonance (ILR). The theory states that the angle of the major-axis of these closed elliptical orbits twists as a function of radius such that orbits align to make $m=2$ spiral arm features. The speed of $\Omega_\mathrm{p}=\Omega-\kappa/2$ is almost constant irrespective of radius for a normal galaxy's rotation curve, which helps maintain an almost stable feature for a long time. These kinetic density-wave spirals with the pattern speed of $\Omega_\mathrm{p}=\Omega-\kappa/2$ are also created from the tidal interaction with another galaxy \citep[e.g.,][]{Antoja+Romas+Lopez-Guitart+22} and also where the self-gravity of stars is not important, such as the outer disk \citep[e.g.,][]{Hu+Sijacki16}.

On the other hand, in N-body simulations of an isolated Milky Way-sized disk galaxy, spiral arm features rotate at the same speed as the stars (the pattern speed being a decreasing function of radius for flat rotation curves), and hence are sometimes referred to as "co-rotating". This leads to individual spiral arms winding up and disappearing within a dynamical time (an epicycle period). However, spiral arm features are always present because although individual spiral arms disappear, new spiral arms form in other places of the disk, and are therefore called transient and recurrent structures \citep{Sellwood11}. This scenario is sometimes called dynamic spiral arm scenario \citep{Dobbs+Baba14}.

In the Milky Way disk, there is mixed observational support for both scenarios. The density-wave scenario predicts that star-forming gas, young stars, and the loci of spiral arms should be spatially offset, owing to the relative velocity difference between spiral arms and gas and stars. For example, from the compilation of various observational studies, \citet{Vallee21} claims that there are offsets between different spiral tracers, e.g. molecular gas and different ages of stars, and suggests a spiral arm pattern speed of $\Omega_\mathrm{sp}=12-17$~km~s$^{-1}$~kpc$^{-1}$. In addition, several studies suggest that the moving groups found in the distribution of the radial and rotation velocities of stars in the solar neighborhood, as well as the ridge-like features in the rotation velocity distribution as a function of the radius (Section~\ref{sec:galseismo}), can be explained by the kinetic density-wave scenario. For example, using test particle simulations (i.e. no self-gravity) with the spiral arm potentials with a rigid-body rotation, \citet{Barros+Perez-Villegas+lepine+20} demonstrate that the phase space features observed in the Gaia data are reproduced with the $m=4$ spiral arms with the pitch angle of $14^{\circ}$ and the pattern speed of $\Omega_\mathrm{sp}=28.5$~km~s$^{-1}$~kpc$^{-1}$.

On the other hand,  \citet{Castro-Ginard+McMillan+Luri+21} find that open clusters with different ages seem to show no offset, and conclude that the pattern speeds of the spiral arms consistent with the rotation speed of stars in the radial range between 5 and 12~kpc. This is more consistent with the dynamic spiral arm scenario. The observed vertical \citep{Asano+Kawata+Fujii+24} and in-plane \citep{Funakoshi+Matsunaga+Kawata+24} kinematics in the spiral arm traced by high-mass star-forming regions is consistent with what is seen in N-body simulations with dynamic arms. Comparing kinematics of Cepheid variables around the spiral arms with N-body simulations, \citet{Funakoshi+Matsunaga+Kawata+24} suggest that the Local and Outer arms are growing, while the Perseus arm is disrupting, which supports the transient, recurrent nature of dynamic spiral arms. 

The tension of these scenarios remains even with the exquisite observational data now available for the Milky Way disk. This is partly because of our location within the Galactic disk, and we cannot see a clear picture of the structure of the disk from the outside. To resolve the tension of these scenarios, we need to more clearly identify the location of the spiral arms in the Milky Way disk, especially with relatively older thin disk stars, which trace the gravitational potential of the arms. Also, the spiral arms in the Milky Way could be flocculent, as suggested by several studies \citep[e.g.,][]{Colombo+Duarte-Cabral+Pettitt+22}. More observational data and comparison with simulation models are required to understand the nature and structure of the spiral arms of our Galaxy.

\begin{figure}[t]
\centering
\includegraphics[width=\textwidth]{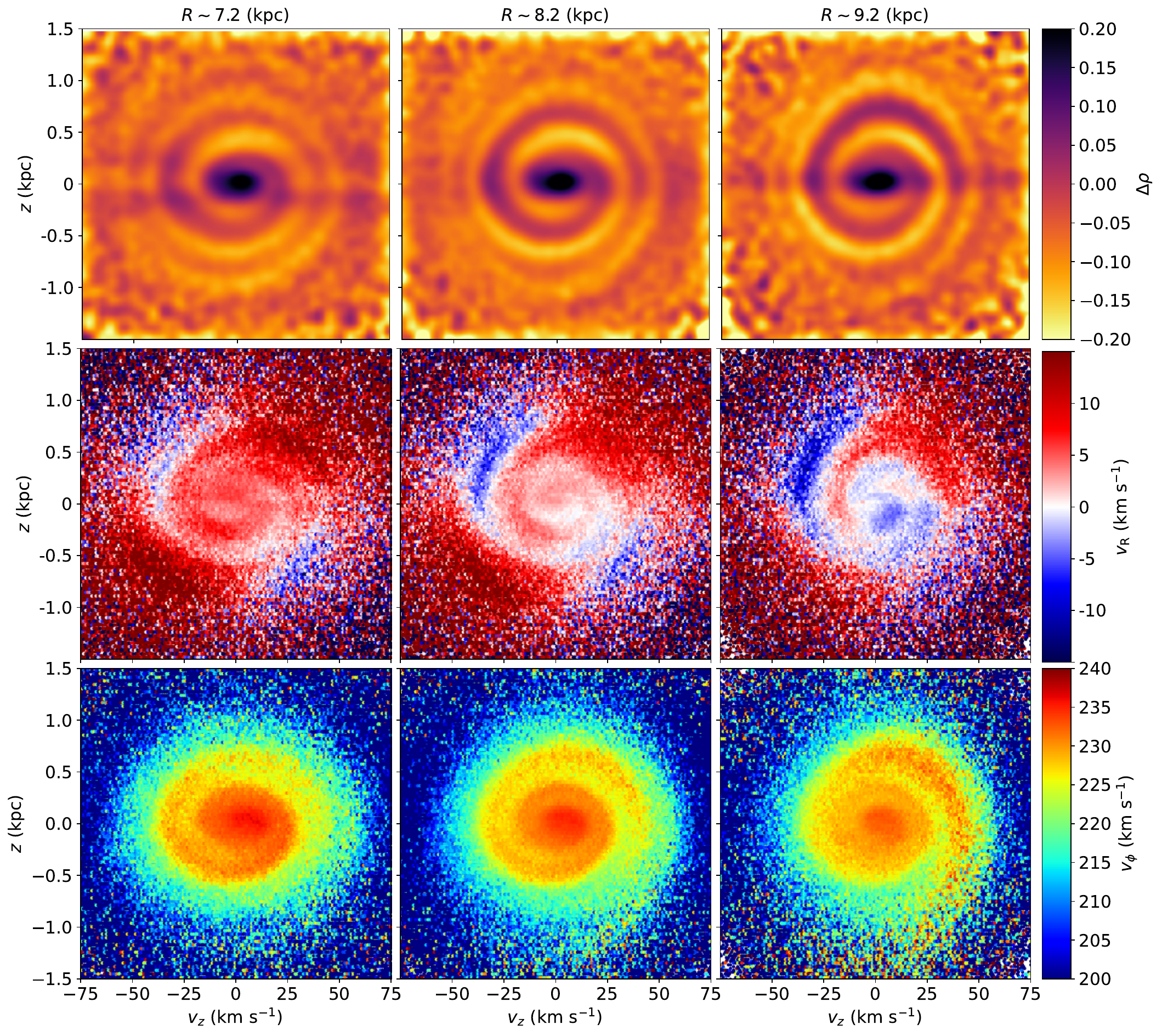}
\caption{`Phase spirals shown in relative density (upper row), as a function of radial velocity, $V_{\mathrm{R}}$ (middle row), and azimuthal velocity, $V_{\phi}$ (lower row) for 1 kpc volumes of Gaia DR3 data centered around $R=7.2$ kpc (left column), $R=8.2$ kpc (i.e. the Solar neighborhood; middle column) and $R=9.2$ kpc (right column); see e.g. \citet{Antoja+Helmi+Romero-Gomez+18,Antoja+Ramos+23}.}
\label{fig:phasespirals}
\end{figure}

\subsubsection{Impacts of the bar and spiral arms: radial migration}
\label{sec:radial-migration}

Non-axisymmetric structures like the bar and spiral arms give rise to "co-rotation resonances": special radii where the bar/spiral arms rotate at the same speed as the stars. 
\citet{Sellwood+Binney02} show that the co-rotation resonance of the spiral arms and bars can induce significant radial migration of stars. This phenomenon occurs when stars lose and/or gain angular momentum as a result of gravitational torques imparted by non-axisymmetric structures, which causes stars to change their guiding center radii: the radius where stars on circular orbits of a given angular momentum would occupy. 
The terminology of "radial migration" or "radial mixing" is used for the change of the angular momentum of stars with or without orbital heating (an increase in orbital eccentricity). \citet{Schoenrich+Binney09a} dubbed the change in the angular momentum as "churning", and the increase in the amplitude of the radial epicycle motion, or orbital heating, as "blurring". 
Radial migration due to a co-rotation resonance as suggested by \citet{Sellwood+Binney02} corresponds to churning without blurring. 

The process of radial migration complicates Galactic archaeology studies because the current angular momentum of some stars will likely be quite different from their angular momentum at birth. \citet{Schoenrich+Binney09a} include churning and blurring in a multi-zone Galactic chemical evolution model and demonstrate that the radial mixing of stars helps to explain the scatter of the age metallicity relation observed for the solar neighborhood stars (Section~\ref{sec:formation}), and also the bimodal distribution of the high-[$\alpha$/Fe] and low-[$\alpha$/Fe], chemically-defined thick and thin disk populations (Figure~\ref{fig:feh_alphafe_rcol}). 

Using numerical simulations of a Milky Way-sized galaxy, \citet{Grand+Kawata+Cropper12} show that because the dynamic spiral arm is co-rotating at all radii, radial migration occurs at every radius. In this case, the stars migrating outward stay on the trailing side of the spiral arm, gaining torque from the arm and staying close to the apo-center phase of their orbit. On the other hand, stars migrating inward stay on the leading side of the spiral arm, losing angular momentum and staying close to peri-center phase. It means that the dynamic spiral arms induce the systematic radial migration around the spiral arms at every radius. Finding such radial migration in action could resolve the tension on the nature of spiral arms (Section~\ref{sec:spirals}). 

\citet{Frankel+Sanders+Ting+20} fit the APOGEE data with an analytical disk galaxy model with parameters describing the inside-out formation of the disk as well as churning and blurring. They assume that stars formed on circular orbits and with a metallicity inherited from their natal star-forming gas; the radial metallicity profile of the gas disk is described with a constant metallicity gradient and an intercept that changes with time. Under these assumptions, each star has a unique metallicity that depends on their formation radius and time. On top of this, their model includes churning and blurring and sets the efficiencies of these processes as fitting parameters. Their best-fit model indicates that churning is more important than blurring in the Milky Way disk, which suggests that radial migration due to the co-rotation resonance is more important than heating. Interestingly, \citet{Okalidis2022} show that cosmological numerical simulations of Milky Way-mass spiral galaxies predict a similar amount of churning to what is reported by \citet{Frankel+Sanders+Ting+20}.

Using the same logic of a star-forming gas disk with a negative metallicity gradient and no scatter, the birth radii of stars can be inferred from their age and metallicity \citep{Minchev+Anders+Recio-Blanco+18}. Because the Sun's metallicity is higher than the stars around us, it is likely that the Sun radially migrated from the inner disk at around $R=5-7$~kpc \citep{Wielen+Fuchs+Dettbarn96}. By analyzing the orbits of the stars migrating from the inner disk to the solar radius in an N-body simulation, \citet{Tsujimoto+Baba20} discuss that if the Sun migrated significantly, during the migration the Sun needed to stay close to the spiral arm for a long time, which may have affected the Earth's environment, and therefore may be associated to historical events like snowball Earth, when all water on the Earth surface was frozen. Hence, the radial migration process of the Sun could be constrained by the geological history of the Earth, and/or the history of the Earth could be closely related to the past journey of the Sun in the Milky Way disk. 

\subsection{Galactoseismology: Phase Spirals/Ridges/Corrugations}
\label{sec:galseismo}

\subsubsection{Phase spirals and ridges}

One of the most striking discoveries of the Gaia data which impacted our view of the Milky Way disk is the fine sub-structure in the phase space distribution of the Milky Way disk stars \citep[e.g.,][]{Antoja+Helmi+Romero-Gomez+18,Kawata+Baba+Ciuca+18}. \citet{Antoja+Helmi+Romero-Gomez+18} find that the distribution of the vertical position of stars relative to the mid-plane of the disk, $z$, and the vertical velocity of stars, $V_\mathrm{z}$, in Gaia DR2 shows a $m=1$ spiral pattern. Figure \ref{fig:phasespirals} shows this spiral pattern in Gaia DR3 data at three different galactic radii, slightly inside the Solar radius (left column), in the Solar neighborhood (center column) and slightly outside the solar radius (right column). The spiral in number density is clearly visible in relative density, $\Delta\rho$, when subtracting a smooth background distribution as shown in the upper row. The feature is also prominent when highlighted with the rotation velocity (lower row), $V_\mathrm{\phi}$, or radial velocity, $V_\mathrm{R}$ (middle row), i.e. the vertical and in-plane kinematics are correlated. 

This feature is now called ``phase spirals''. With data from Gaia DR3, the phase spiral has been mapped across a large region of the Galactic disk in both physical space \citep[e.g.][]{Antoja+Ramos+23}, and as a function of orbital parameters \citep[e.g.][]{Hunt+Price-Whelan+Johnston+22}. Distinct $m=2$ arm phase spirals have been discovered in the inner disk \citep{Hunt+Price-Whelan+Johnston+22}, and equivalent phase spirals are seen in the radial direction $\Delta R-V_{\mathrm{R}}$ \citep{Hunt+Price-Whelan+Johnston+24}. All these phase spirals are the signature of phase mixing following some perturbation to the disk, the origin of which is discussed in Section \ref{sec:perturbation_origin} \citep[and see][for a thorough review]{Hunt+Vasiliev24}.

\begin{figure}[t]
\centering
\includegraphics[width=\textwidth]{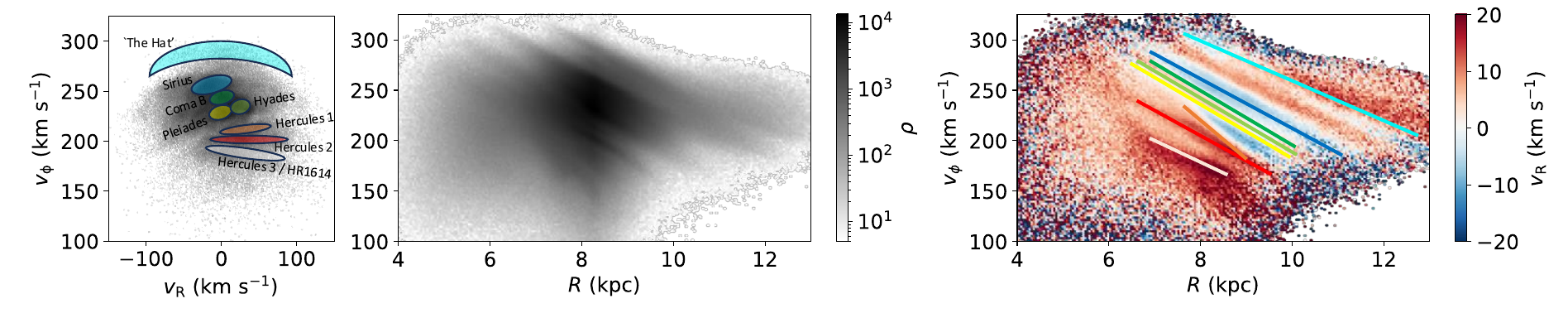}
\caption{\textbf{Left:} $V_{\mathrm{R}}-V_\phi$ plane with the classical moving groups marked. \textbf{Center:} $R-V_\phi$ plane in number density. \textbf{Right:} $R-V_\phi$ plane as a function of radial velocity, $V_{\mathrm{R}}$, with the rough location of the moving groups marked, matching the colors from the left panel \citep[see e.g.][]{Hunt+Bub+Bovy19}.}
\label{fig:rvphi}
\end{figure}

\citet{Kawata+Baba+Ciuca+18} and \citet{Antoja+Helmi+Romero-Gomez+18} show that there are many ridge-like features in the stellar distribution of the rotation velocity, $V_\mathrm{\phi}$, as a function of radius, $R$. They are now known to be the radial extension of the moving groups first identified by Olin Eggen in the $V_\mathrm{\phi}-V_\mathrm{R}$ phase space plane for the solar neighborhood stars \citep[see][and references therein]{Eggen96}. \citet{Fragkoudi+Katz+Trick+19} show that the ridge-like features in the $R-V_\mathrm{\phi}$ plane are correlated with $V_\mathrm{R}$ as well. Figure \ref{fig:rvphi} shows the classic $V_{\mathrm{R}}-V_{\phi}$ plane with a rough illustration of the moving groups (left). The middle panel shows the $R-V_\phi$ plane in number density, where it can be seen that these diagonal ridge-like features stretch over several kpc in the disc. The right panel shows the $R-V_\phi$ plane as a function of $V_{\mathrm{R}}$, with the rough location of the moving groups marked, with colors that match the groups in the left panel.

\subsubsection{Corrugations and warp}

Using Gaia DR1 \citet{Schoenrich+Dehnen18} revealed a wave-like feature in $V_\mathrm{z}$ as a function of the angular momentum of the stars. \citet{Friske+Schoenrich19} build on this with Gaia DR2 data, finding the wave-like feature is also present in the radial velocites, $v_{\mathrm{R}}$ as a function of angular momentum. The vertical, $V_\mathrm{z}$, and in-plane radial, $V_\mathrm{R}$, velocities are likely related to each other and due to a common origin and/or one induces the other, as discussed below. 

Cepheid variables are young (a few 100 Myr old) variable stars, and are known to be a standard candle to provide precise distances, because their absolute magnitude is correlated to the period of their flux variations. Hence, the distances to Cepheids can be measured more precisely than the parallax measurement of Gaia especially at a larger distance ($d>\sim5$~kpc).
\citet{Skowron+Skowron+Mroz+19} map the spatial distribution of about 2000 Cepheids in the Milky Way disk and reveal clearly the structure of the warp in the Galatic outer disk up to about $R=20$~kpc. We define $\phi=0$ as the Sun--Galactic center line of the Galactocentric azimuthal angle, $\phi$, and $\phi$ to be positive in the direction of Galactic rotation, as shown in Figure~\ref{fig:mwd-image}. At $R>\sim10$~kpc, the disk bends downward behind the direction of rotation of the Sun, i.e. at $\phi<0$, while at positive $\phi$ it goes upward. The observed amplitude is about $\Delta |z|\sim1$~kpc at $R\sim15$~kpc. Hence, the line of nodes of the warp is close to the Sun--Galactic center line, but it is slightly on the positive $\phi$ side. \citet{Chen+Wang+Deng+19} suggest that the line of nodes of the warp is twisted, i.e. it has a different angle at different radii, which is a sign of the precession of the warp. Using Gaia DR2 data and analyzing the position and kinematics of about 12 million giant stars, \citet{Poggio+Drimmel+Andrae+20} measure the precession of the warp to be about 11~km~s$^{-1}$~kpc$^{-1}$, where the positive value means the warp is precessing in the same direction as the rotation of the disk. However, the exact shape of the line of nodes and the speed of the precession are hotly debated, and our view of the outer edge of the Galactic disk requires further exploration. 

Furthermore, \citet{Poggio+Khanna+Drimmel+24} subtract the general trend of the disk bending due to the warp and analyse the remaining structures present in the Gaia data. They find the vertical corrugation in the young ($<$a few 100~Myr) disk stars is propagating outward. They suggest that this vertical corrugation reaches up to $|z|\sim150-200$~pc and has a radial length of about 3~kpc. The corrugation is seen up to $R\sim20$~kpc when traced by the Cepheids. 

This vertical corrugation is also found in the star-forming gas filaments. \citet{Alves+Zucker+Goodman+20} have discovered a filamentary gas structure, which shows a vertical wave-like feature with a vertical amplitude of about 160~pc and wavelength of 2~kpc, around the Sun, called the Radcliffe wave. This wave spans about 3 kpc along the inside of the Local arm. The wavelength in the radial direction is about 1~kpc, which is significantly shorter than the corrugation seen in the young stars in \citet{Poggio+Drimmel+Andrae+20}. It is not yet clear if or not they are related to each other. 

The older stars in the disk seem to show even more spectacular vertical wave-like features. In the anti-center direction, several over-density structures are observed and spread to high Galactic latitudes up to about $|b|=30^{\circ}$. \citet{Bergmemann+Sesar+Cohen+18} take high-resolution spectra for the red giant branch stars associated with two of these structures, called the Triangulum–Andromeda (TriAnd) and A13 overdensities, and obtain their chemical compositions. From the similarity of their abundances to the Galactic thin disk, they conclude that both TriAnd and A13 are part of the outer thin disk. The A13 structure is at around $R=16$~kpc, and TriAnd is around $R=28$~kpc. Also, A13 is above the mid-plane of the inner disk by about $z=5$~kpc, while TriAnd is below by 5~kpc. This is a much bigger amplitude than the warp seen in the younger stars. In addition, TriAnd is located at $\phi\sim30^{\circ}$, where the warp bends upward, and A13 is around $\phi=0^{\circ}$, which is close to the line of nodes of the warp. Hence, the older stars seem not to follow the warp, but have their own vertical corrugation mode in the outer disk.

\begin{figure}
\centering
\includegraphics[width=\textwidth]{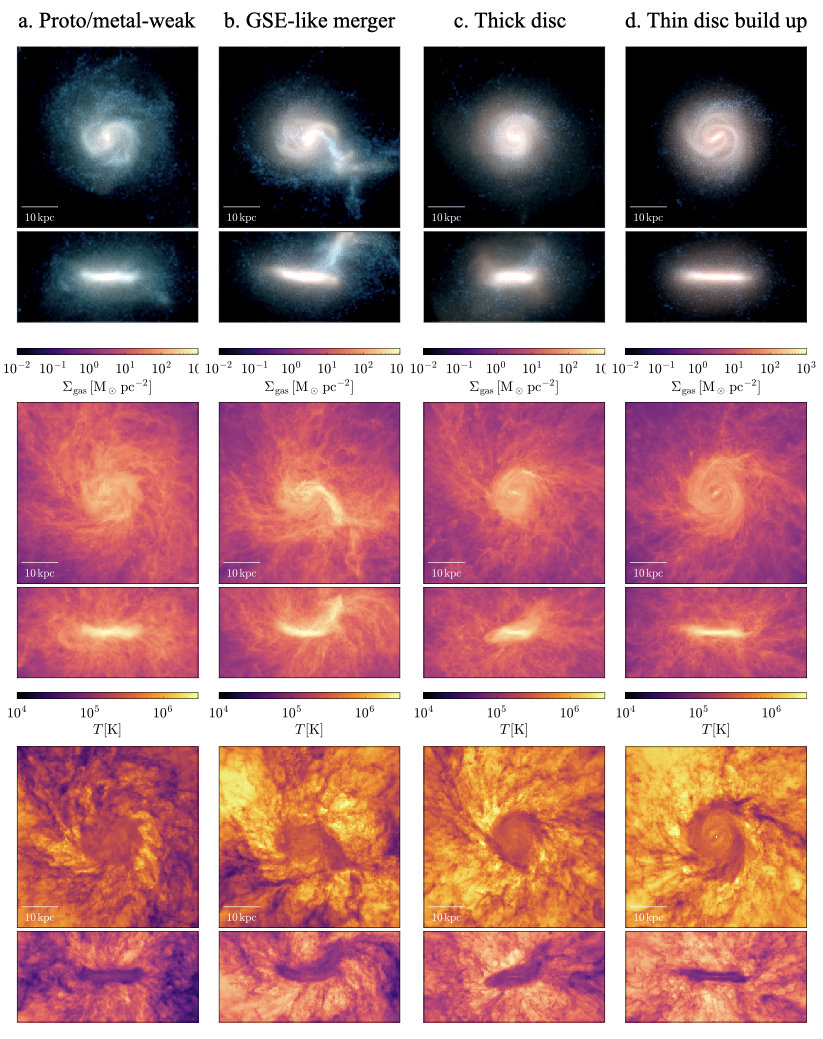}
\caption{Simulation snapshots depicting important stages of disk formation. The top-, middle-, and bottom-rows show the the face-on and edge-on view of stellar light (blue and red indicate young and old stars, repectively), gas density, and gas temperature at each stage, respectively. The highlighted stages are: a) the metal-weak proto-disk at early epochs; b) the GSE merger event; c) the thick disk formation; and d) the build-up of the thin disk. }
\label{fig:AurigaSS}
\end{figure}

\subsubsection{The sources of the perturbation}\label{sec:perturbation_origin}

A number of theoretical studies and numerical simulation studies have been published since Gaia uncovered the velocity substructures in the Galactic disk. \citet{Antoja+Helmi+Romero-Gomez+18} suggest that the phase spiral they find in the Gaia data arises from phase mixing from an out-of-equilibrium condition caused by some perturbation. From the wraps of the phase spiral, they suggest that the impact was $300-900$~Myr ago, which is a similar time to the expected last passage of the Sagittarius dwarf galaxy. It is also demonstrated that a Sagittarius dwarf-like perturber can produce phase spiral-like features \citep[e.g.][]{Laporte+Minchev+Johnston+19}. However, the required mass of the perturber to have sufficient impact on the disk is higher than the currently estimated mass of the Sagittarius dwarf. Although several simulations are able to qualitatively reproduce such phase spirals, none have yet reproduced the specific configuration (amplitude, pitch angle, and phase as a function of Galactic position) seen in Gaia from a Sagittarius dwarf-like simulation \citep[e.g.][and references therein]{Bennett+Bovy+Hunt22}.

There are alternative mechanisms that can create phase spirals. As explained above, single-armed phase spirals can be created by the displacement of the phase space distribution of the stars in $z-V_\mathrm{z}$ phase space. Hence, any vertical corrugation can trigger such $m=1$ phase spirals. Using N-body simulations, \citet{Khoperskov+DiMatteo+Gerhard+19} demonstrate that the bar buckling, which may occur when the BPX-bulge is created in the Milky Way disk, can trigger such a vertical wave and generate the phase spirals seen in the disk. However, BPX-bulges can also form through resonant growth without undergoing a bucking instability, and the nature of the Milky Way's BPX-bulge is still debated \citep[see e.g.][for an example of the BPX-bulge formation without buckling]{Baba+Kawata+Schoenrich22}. 

Alternatively, the phase spirals can be created by interaction with dark matter. \cite{Tremaine+Frankel+Bovy+23} show that stochastic bombardment of the disk by dark matter subhalos can cause long-lived phase spirals. \citet{Grand+Pakmor+Fragkoudi+23} show that earlier interactions of the Milky Way with the Sagittarius dwarf galaxy when it was much heavier can induce a significant wake in the Milky Way's dark matter halo which lasts for several Gyr. This wake can then continue to perturb the disk long after the initial interaction, potentially creating and maintaining the phase spirals until the present day even after Sagittarius has lost most of its mass.

Regardless of the origin of the perturbation, the perturbation(s) likely caused both the corrugations and the phase spirals, and possibly the warp. As mentioned above, the phase spirals are well correlated with the in-plane motion, such as the radial and azimuthal motion of the stars. The vertical corrugation and spiral arms in the disk could also be related to each other. In fact, a tidal interaction, such as the impact of the Sagittarius dwarf, can trigger both vertical corrugations and in-plane spiral arm features, as demonstrated by \citet{Bland-Hawthron+Tepper-Garcia21}. In fact, spiral arms can also trigger vertical breathing motion, i.e. expansion and compression motion \citep{Debattista14,Asano+Kawata+Fujii+24}. Such breathing motion can induce two armed phase spirals, such as those found by \citet{Hunt+Price-Whelan+Johnston+22}. It is likely that the vertical motion and in-plane motions are coupled to each other, and modelling both together is likely key to breaking the degeneracies between the various models. 

\begin{figure}[t]
\centering
\includegraphics[width=0.75\textwidth]{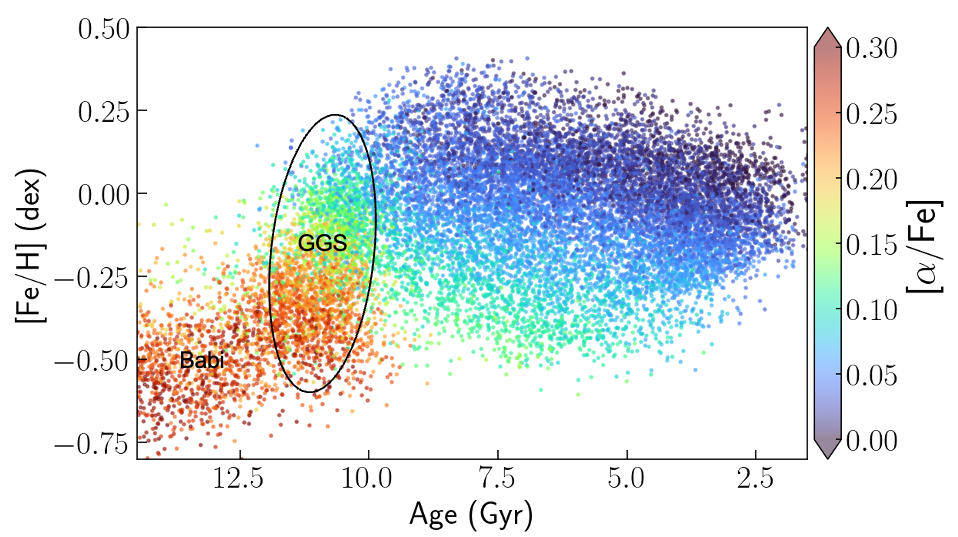}
\caption{The age-metallicity relationship of APOGEE DR17 giants (the same sample as Figure~\ref{fig:feh_alphafe_rcol}). The colour of the dots indicates [$\alpha$/Fe]. We highlight the older population Babi and the GGS (see text).}
\label{fig:AMR-GGS}
\end{figure}

\section{The formation of the Milky Way disk: Proto-disk, Thick disk, and Thin disk}
\label{sec:formation}

The recent advance of computing power and significant improvements in modeling the physics of galaxy formation have enabled computer simulations to follow the formation and evolution process of Milky Way-like galaxies from very high redshifts (e.g. $z\sim 100$) until the present-day. These simulations demonstrate that, under the current standard $\Lambda$-dominated cold dark matter ($\Lambda$CDM) cosmogony, Milky Way-like disk galaxies inevitably experience chaotic mergers at early epochs of formation. To build a Milky Way-like disk galaxy with a prominent thin disk, this period should be  followed by a longer term ($\sim8-10$~Gyr) quiet phase characterized by the gradual build-up of the thin disk in the absence of major mergers \citep[although minor mergers and galactic interactions can occur;][]{Brook+Kawata+Gibson+Freeman04,Grand+Bustamante+Gomez18}. This thin disk is where the majority of the Milky Way's stars now reside. Although simulations also show that a thinner disk can form at early epochs, the prevalence of merger events ensures that such a disk would sustain repeated dynamical impacts that scatter stars onto orbits that would now be observed as part of the thick disk and stellar halo.

In the Gaia data, there is evidence that the early epoch of merger activity concluded around $8-10$~Gyr ago with a significant merger known as Gaia-Sausage-Enceladus \citep[hereafter GSE,][]{Belokurov+Erkal+Evans+18,Helmi+Babusiaux+Koppelman+18}, the stellar debris of which makes up about half of the stellar halo. Interestingly, signs of satellite accretion in the stellar halo have been claimed with the ESA's Hipparcos data. \citet{Brook+Kawata+Gibson+Flynn03} note that the combination of the Hipparcos data and the line-of-sight velocity and metallicity information from spectroscopic surveys shows a group of stars with low metallicity on high-eccentricity orbits, as seen also in the historical \citet{Eggen+Lynden-Bell+Sandage62} work. However, rather than regarding this group of stars as evidence of the rapid collapse of the Milky Way halo as considered in \citet{Eggen+Lynden-Bell+Sandage62}, \citet{Brook+Kawata+Gibson+Flynn03} used comparisons with their numerical simulations to demonstrate that the phase space distribution of this group of stars is more naturally explained by the remnants of a dwarf galaxy accreted into the Milky Way in the past. Now, it is clear that the phase space distribution of the group of stars they found is consistent with that of the GSE stars found in the Gaia data. 
Aside from clear differences in their orbital characteristics, many studies \citep[e.g.,][]{Haywood+DiMatteo+Lehnert18,Helmi+Babusiaux+Koppelman+18} show also that the chemical abundances of GSE stars are different from those that are likely "in-situ" stars, i.e. stars formed within the main progenitor of the Milky Way. These findings agree with earlier results found prior to Gaia by \citet{Nissen+Schuster10}.

The Gaia data show also that the GSE merger has a significant impact on the "in-situ" stars that were already formed at the time of the merger: these stars were "splashed" out of the disk and into halo, and are currently observed as relatively metal-rich halo stars compared to the GSE remnants. This component of the stellar halo was identified by \citet{DiMatteo+Haywood+Lehnert+19} and \citet{Gallart+Bernard+Brook+19}, and was dubbed the "Splash" by \citet{Belokurov+Sanders+Fattahi+20}. This means that the Galactic disk was already building up well before the GSE merger. This proto-Galactic disk is claimed to be identified by \citet{Belokurov+Kravtsov22}: they found that the median rotation velocity of metal-poor stars increases sharply, or is "spun-up", in the metallicity range [Fe/H]$\sim-1.5$ to [Fe/H]$\sim-0.9$. They call this component "Aurora". Similar trends are found in results summarised by various metaphors such as the "boiling hot" phase and the "poor old heart" by \citet{Conroy+Weinberg+Naidu+22} and \citet{Rix+Chandra+Andrae+22}, respectively. Interestingly, these new studies are consistent with what is found in the Hipparcos data of the solar neighborhood stars in \citet{Chiba+Beers00}, who established the clear increase of the mean rotation velocity of stars with metallicity at [Fe/H]$>\sim-1.7$. These studies show also that this metal-poor spin-up component is old and centrally concentrated. Hence, this is likely to be the starting phase of the Milky Way disk, which unsurprisingly began from a smaller, metal-poor disk. 

Proto-disk galaxies such as these likely experienced many mergers. Because they are still too small to establish the hot gas halo, their mergers are likely to be gas-rich mergers. Thanks to the dissipative nature of gas, gas-rich mergers lead to a disk-like galaxy \citep{Robertson+Bullock+Cox+06,Brook+Richard+Kawata+07}, though it may not be as thin as the disk galaxies of today. In addition, as suggested by \citet{Noguchi98} with his numerical simulations, thermal instability is quite efficient in gas-rich disks at early epochs, and this likely generates large star-forming clumps that continuously puff-up the disk through dynamical scattering. As mentioned above, the GSE merger splashed some proto-disk stars onto eccentric, halo-like orbits. However it is likely that some fraction of proto-disk stars remain in what is now regarded as the thick disk \citep{Grand+Kawata+Belokurov+20}.

The GSE merger is now considered to be the last significant galaxy merger for the Milky Way. It follows that the Milky Way has had about $8-10$ Gyr of evolution during which no major merger has occurred. In such a quiet period, the galaxy is thought to have built a thin disk via the smooth accretion of halo gas; gas with higher angular momentum falls onto the disk at later epochs \citep{Brook+Governato+Roskar+11} and, as such, builds the thin disk in an inside-out fashion \citep{Brook+Kawata+Gibson+Freeman04}. The Milky Way's current mass is just above the mass threshold required to develop the hot halo gas, i.e. "hot-mode" \citep{Keres+Katz+Weinberg+Dave05}, which cuts-off the direct cold filamentary accretion onto the disk by shock-heating cold gas flows. However, the Milky Way has not always been this massive and must have been under this mass threshold at earlier epochs during which cold flows are not shock heated by the halo gas and instead accrete unimpeded directly onto the disk. \citet{Noguchi18} suggest that the transition from thick disk formation to thin disk formation is because of the transition from cold mode accretion to hot mode accretion as the Milky Way surpasses the mass threshold. 

Using self-consistent cosmological (magneto-)hydrodynamical simulation of a Milky Way-sized galaxy, \citet{Grand+Bustamante+Gomez18} showed that it is possible for such a transition to happen for Milky Way-sized galaxies. Further,  \citet{Grand+Kawata+Belokurov+20} demonstrated that the GSE-like merger can trigger this transition by boosting the total mass of the Milky Way, thus aiding thin disk formation from subsequent smooth, hot-mode gas accretion. Figure~\ref{fig:AurigaSS} shows a sequence of snapshots depicting the stages of disk formation described above, taken from the Auriga-18 cosmological simulation. The top-, middle-, and bottom-rows show the the face-on and edge-on view of stellar light, gas density, and gas temperature at each stage, respectively. At the earliest epoch (column a), the relatively small proto-disk/metal-weak disk forms in the presence of minor mergers and cold-mode gas accretion. Column b) shows the gas-rich GSE merger event, which both scatters stars into the thick disk and "Splash" halo components and induces a starburst by compressing gas into a dense, compact form that contributes to a significant part of the thick disk (column c). Coincident with the finalization of these processes is the increase of halo gas temperature, which marks a transition to hot-mode accretion and the growth of the thin disk (column d). Like many other studies \citep[e.g.,][]{Brook+Kawata+Gibson+Freeman04,Bird+Kazantzidis+Weinberg+13,Agertz+Renaud+Feltzing+21}, these simulations show that the gas disk becomes thinner with time, and thus the thin disk forms "upside-down" as well as inside-out. Interestingly, they show also that the gas and stars belonging to the thin disk evolve with a flared scale-height profile qualitatively similar to the Milky Way thin disk as described in Section~\ref{sec:thin-thick}.

The predicted impact of the GSE merger and subsequent thin disk formation are also qualitatively consistent with the precise chemical abundance measurements from the APOGEE survey and the kinematics from the Gaia data \citep[see][for a review]{Deason+Belokurov24}. Advanced data analysis techniques 
now allow us to infer reliable stellar ages, which is crucial information for the archaeological study of the Milky Way's formation and evolution history. \citet{Ciuca+Kawata+Ting+24} analyse the age-metallicity relation of APOGEE giant stars (Figure~\ref{fig:AMR-GGS}), and identify three phases of the Milky Way formation: the metal-poor proto-disk phase, which they called "Babi"; a rapid increase in metallicity in a short time, which they dubbed the "Great Galactic Starburst (GGS)"; followed by the metal-rich thin disk formation phase. Note that the Babi corresponds to the late stage of the spin-up phase (like Aurora) because Figure~\ref{fig:AMR-GGS} focuses on the stars after the proto-disk is established at [Fe/H]$>-1$. 
The subsequent rapid increase of metallicity in the GGS phase supports the idea that the GSE merger was a gas-rich galaxy merger, which would trigger a starburst and in turn spur the rapid chemical enrichment of gas and successive generations of stars. The lowest metallicity component, i.e. the beginning of GGS, also shows high [$\alpha$/Fe]. This also indicates the rapid increase of SNe~II that naturally follow from an increase in star formation rate as a result of the gas-rich merger. 


\begin{figure}[t]
\centering
\includegraphics[width=\textwidth]{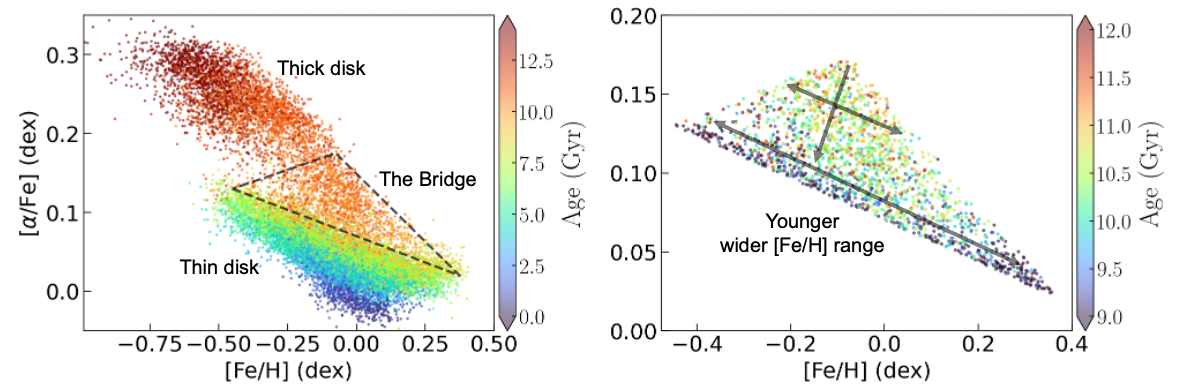}
\caption{The distribution of [$\alpha$/Fe] and [Fe/H] colored by age for stars in Figure~\ref{fig:feh_alphafe_rcol}. The dashed triangle region in the left-hand panel is referred to as the Bridge, and is a transition region between the thick (old high-[$\alpha$/Fe] population) and thin (young low-[$\alpha$/Fe] population) disks. An age gradient is apparent, as indicated by the near vertical downward arrow, in a close-up of the Bridge region, shown in the right-hand panel, and the range of [Fe/H] becomes wider, as indicated by the double arrows for the younger stars in the Bridge region.}
\label{fig:fehalfe-age}
\end{figure}

\begin{figure}[t]
\centering
\includegraphics[width=\textwidth]{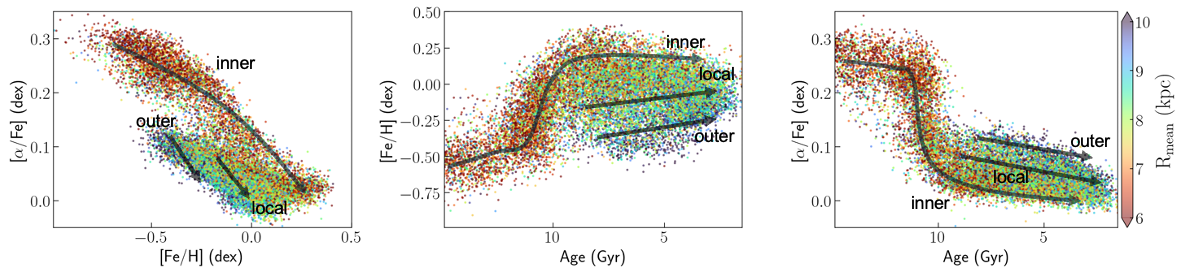}
\caption{The distribution in [$\alpha$/Fe] versus [Fe/H] (left-hand panel), [Fe/H] versus age (middle panel), and [$\alpha$/Fe] versus age (right-hand panel) coloured by mean orbital radius, $R_\mathrm{m}$. The ‘inner’, ‘local’, and ‘outer’ arrows indicate the schematic chemical evolution paths at the inner ($R_\mathrm{m}\sim6$ kpc), local, i.e. solar radius ($R_\mathrm{m}\sim8$ kpc), and outer discs ($R_\mathrm{m}\sim10$ kpc), respectively. The metal-poor, outer disk stars follow a different chemical evolution pathway than the inner disc. These evolutionary paths are shown to describe qualitative trends of the chemical evolution at the different radii of the Galactic disk, and are not meant to indicate the chemical evolution paths quantitatively. Adapted from \citet{Ciuca+Kawata+Miglio+21}.}
\label{fig:fehalfe-agefeh-agealfe}
\end{figure}

From the age-[Fe/H]-[$\alpha$/Fe] relation of the APOGEE giants data, \citet{Ciuca+Kawata+Miglio+21} suggest that there is a transition phase, which they called "the Bridge", between the high-[$\alpha$/Fe] thick disk and the low-[$\alpha$/Fe] thin disk populations (Figure~\ref{fig:fehalfe-age}). Such transition population of stars are also noted in the earlier studies, such as \citet{Adibekyan+Figueira+Santos+13}. \citet{Ciuca+Kawata+Miglio+21} show that, in the Bridge, [$\alpha$/Fe] decreases with age, and the width of the range of [Fe/H] at fixed age increases also. Their interpretation is that, during the high-[$\alpha$/Fe] thick disk formation phase before the Bridge, the prevalence of gas-rich mergers and cold-mode accretion drive a high degree of turbulence that leads to a well-mixed gas disk. After the GSE merger settled, which corresponds to the beginning of the Bridge, the thin disk grows in mass and size such that a radial metallicity gradient (with higher metallicity in the inner disk) develops, which leads to a wider range of metallicities for younger thin disk stars. As mentioned above, although the star-forming gas disk could have a clear metallicity gradient with no metallicity variation at a fixed radius, radial migration brings stars formed in both the inner and outer disk to the solar neighborhood. Hence, we can observe a wide variety of metallicity of both young and old thin disk stars \citep{Schoenrich+Binney09a}. The transition from the thick disk phase, which, we consider, happened only in the inner disk, to the thin disk phase are summarized in the observed [Fe/H]-[$\alpha$/Fe] relation, age-[Fe/H] relation, and age-[$\alpha$/Fe] relation in Figure~\ref{fig:fehalfe-agefeh-agealfe}. The schematic chemical evolution pathways are indicated in the figure. Note that the absolute value of the age used in Figures~\ref{fig:AMR-GGS}-\ref{fig:fehalfe-agefeh-agealfe} is less meaningful, but the relative difference of the ages is precise and meaningful. Although the stars should not be older than the age of the Universe, the prior of the age of the Universe is not used, because the prior of the old age limit forces the inferred age of the old stars to be similar. Hence, the inferred ages are allowed to exceed the age of the Universe. This helps to make the relative age difference between the old stars more clear. 

Similar scenario to arrows in Figure~\ref{fig:fehalfe-agefeh-agealfe} is suggested by \citet{Haywood+Snaith+Lhnert+19}, although they consider that the outer thin disk formed earlier than the inner thin disk population. Note also that the arrows in Figure~\ref{fig:fehalfe-agefeh-agealfe} are only schematic trends merely reflecting the authors' current view. More data with the accurate age measurements are crucial to fully understand how the transition from the thick disk to thin disk formation happened, how the GSE impacted on it, and if the bar formation, which is likely around similar time (Section~\ref{sec:bar-age}), is also related to this transition \citep[e.g.,][]{Merrow+Grand+Fragkoudi+24}.

\section{Summary and Future Prospects}
\label{sec:summary}

The Gaia mission and both photometric and spectroscopic surveys of the Galactic stars have revolutionised our view of the Milky Way disk. The Milky Way disk is no longer considered to be a quiet stable disk galaxy, but is rather an out-of-equilibrium structure reeling from perturbation(s) caused by accreting satellites and/or dark matter halo. This leads to many challenges for us to accurately measure the mass and size of the Milky Way disk, and use the Galactic disk stars to infer the density of the dark matter. However, a challenge is always an opportunity. The data from Gaia and complementary surveys keep improving and our understanding of the phase space sub-structures is advancing. More precise measurements of these motions and comparisons with dynamical models and numerical simulation must allow us to use such seismic information of the Milky Way disk to infer both the nature of the waves, including the bar and spiral arms, and the composition of the Galaxy, including the density profiles of the stellar disk and dark matter. 

In addition to the upcoming Gaia~DR4 with five years of the Gaia mission data and the final data release of more than ten years of data, further advances of observational data are anticipated. The Prime Focus Infrared Microlensing Experiment (PRIME)\footnote{\url{http://www-ir.ess.sci.osaka-u.ac.jp/prime/index.html}}, a wide field 1.8~m telescope with a NIR camera at the South African Astronomical Observatory, started the inner Milky Way region survey. The Vera C. Rubin Observatory's the Legacy Survey of Space and Time (LSST)\footnote{\url{https://rubinobservatory.org/}} is planned to begin in 2025, and will provide time series photometric data in the large region of the Southern hemisphere sky, including the Galactic bulge/bar region. Ground-based spectroscopic surveys with the next generation multi-object spectrograph in the Southern hemisphere, such as optical 4-meter Multi-Object Spectrograph Telescope (4MOST)\footnote{\url{https://www.4most.eu/cms/home/}} on the Visible and Infrared Survey Telescope for Astoronomy (VISTA) telescope and NIR Multi-Object Optical and Near-infrared Spectrograph (MOONS)\footnote{\url{https://vltmoons.org/}} on VLT will further add value to these data and will help us understand the kinematics and stellar abundance distributions of the inner Milky Way disk. In the Northern hemisphere, WHT Enhanced Area Velocity Explorer (WEAVE)\footnote{\url{https://weave-project.atlassian.net/wiki/spaces/WEAVE/overview}} on the William Herschel Telescope (WHT) and Prime Focus Spectrograph (PFS)\footnote{\url{https://pfs.ipmu.jp/}} on Subaru will start science surveys soon, and will provide valuable data for the outer disk and halo stars. These data will help understand the holistic view of the Milky Way's disk formation and perturbation histories and how they are related to the subtle features in the halo stars. 

As described in this chapter, stars are crucial tracers and fossil records to tell us the formation history of the Galaxy. Hence, understanding stellar physics is critical for Galactic astronomy and a wide range of astronomy research. Asteroseismology is now a critical tool for understanding stellar physics. The proposed space mission, High-precision AsteroseismologY in DeNse stellar fields \citep[HAYDN,][]{Miglio+Girardi+Grundhal+21}\footnote{\url{https://www.asterochronometry.eu/haydn/}}, plans to measure the precise time-series photometry of stars in star clusters with homogeneous age and abundances. This will finally allow us to calibrate stellar physics models and will help us to accurately measure the ages of stars. 

Because Gaia is an optical survey, astrometry of stars found in line-of-sights toward the Galactic center is limited to within a few kpc from the Sun. The Japan Astrometry Satellite Mission for INfrared Exploration \citep[JASMINE,][]{Kawata+Kawahara+Gouda+24}\footnote{\url{http://jasmine.nao.ac.jp/index-en.html}} NIR space astrometry mission by ISAS/JAXA will provide precise astrometry for the stars between the Galactic center and the Sun, although their survey region is focused on only the Galactic center region ($<\sim1.5^{\circ}$). Ultimately, the ESA's future planned L-mission, GaiaNIR\footnote{\url{https://www.astro.lu.se/GaiaNIR}} \citep{Hobbs+Brown+Hog+21}, will provide NIR all-sky astrometry of similar precision and depth to Gaia, but in NIR. The GaiaNIR mission is expected to be able to measure astrometry for about eight billion stars, including the stars in the Galatic disk mid-plane and the stars around the super-massive black hole in the center of the Milky Way. In addition, the next-generation of radio telescopes, such as Square Kilometer Array (SKA)\footnote{\url{https://www.skao.int/en}} and Next Generation Very Large Array (ngVLA)\footnote{\url{https://ngvla.nrao.edu/}}, will also provide information on the gas, star formation, and magnetic fields. These multi-wave future surveys of the Milky Way will help uncover the formation process of the Milky Way disk - where our Sun was born and where we live.

\begin{ack}[Acknowledgments]
\ This work is a part of MWGaiaDN, a Horizon Europe Marie Sk\l{}odowska-Curie Actions Doctoral Network funded under grant agreement no. 101072454 and also funded by UK Research and Innovation (EP/X031756/1). This work was also partly supported by the UK's Science \& Technology Facilities Council (STFC grant ST/S000216/1, ST/W001136/1). RJJG is supported by an STFC Ernest Rutherford Fellowship (ST/W003643/1).

This work has made use of data from the European Space Agency (ESA) mission {\it Gaia} (\url{https://www.cosmos.esa.int/gaia}), processed by the {\it Gaia}
Data Processing and Analysis Consortium (DPAC, \url{https://www.cosmos.esa.int/web/gaia/dpac/consortium}). Funding for the DPAC has been provided by national institutions, in particular the institutions participating in the {\it Gaia} Multilateral Agreement.

Funding for the Sloan Digital Sky Survey IV has been provided by the Alfred P. Sloan Foundation, the U.S. Department of Energy Office of Science, and the Participating Institutions. SDSS acknowledges support and resources from the Center for High-Performance Computing at the University of Utah. The SDSS web site is www.sdss4.org. SDSS is managed by the Astrophysical Research Consortium for the Participating Institutions of the SDSS Collaboration including the Brazilian Participation Group, the Carnegie Institution for Science, Carnegie Mellon University, Center for Astrophysics | Harvard \& Smithsonian (CfA), the Chilean Participation Group, the French Participation Group, Instituto de Astrofísica de Canarias, The Johns Hopkins University, Kavli Institute for the Physics and Mathematics of the Universe (IPMU) / University of Tokyo, the Korean Participation Group, Lawrence Berkeley National Laboratory, Leibniz Institut für Astrophysik Potsdam (AIP), Max-Planck-Institut f\"ur Astronomie (MPIA Heidelberg), Max-Planck-Institut für Astrophysik (MPA Garching), Max-Planck-Institut f\"ur Extraterrestrische Physik (MPE), National Astronomical Observatories of China, New Mexico State University, New York University, University of Notre Dame, Observatório Nacional / MCTI, The Ohio State University, Pennsylvania State University, Shanghai Astronomical Observatory, United Kingdom Participation Group, Universidad Nacional Autónoma de México, University of Arizona, University of Colorado Boulder, University of Oxford, University of Portsmouth, University of Utah, University of Virginia, University of Washington, University of Wisconsin, Vanderbilt University, and Yale University.
\end{ack}


\bibliographystyle{Harvard}
\bibliography{reference}

\end{document}